\documentclass{article}

\usepackage{arxiv}

\usepackage[utf8]{inputenc} 
\usepackage[T1]{fontenc}    
\usepackage{url}            
\usepackage{booktabs}       
\usepackage{amsfonts}       
\usepackage{nicefrac}       
\usepackage{microtype}      
\usepackage{lipsum}         
\usepackage{graphicx}
\usepackage{doi}
\usepackage{amssymb}
\usepackage{amsmath}
\usepackage{bm}  
\usepackage{algorithm}
\usepackage{algpseudocode}
\usepackage{multirow}
\usepackage{siunitx} 
\usepackage{adjustbox}
\usepackage{xcolor}   
\usepackage{subcaption}  
\usepackage{float}       
\usepackage{rotating} 
\usepackage{tabularx} 
\usepackage[capitalize]{cleveref}

\usepackage{hyperref}
\hypersetup{
colorlinks=true,
linkcolor=cyan,
filecolor=green,
urlcolor=blue,
citecolor=blue,
}

\title{Coverage-Guaranteed Speech Emotion Recognition via Calibrated Uncertainty-Adaptive Prediction Sets}

\author{ 
	{\includegraphics[scale=0.06]{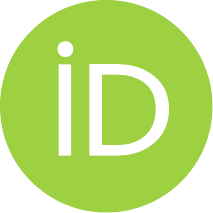}\hspace{1mm}Zijun Jia}\\
    Beihang University\\
	\texttt{21375166@buaa.edu.cn} \\
	\And
	{\includegraphics[scale=0.06]{orcid.pdf}\hspace{1mm}Jinsong Yu*}\\
    Beihang University\\
	\texttt{yujs@buaa.edu.cn} \\
      \And
        {\includegraphics[scale=0.06]{orcid.pdf}\hspace{1mm}Hongyu Long*}\\
         Chongqing University of Posts and Telecommunications\\
	\texttt{longhy@cqupt.edu.cn}\\
         \And
	{\includegraphics[scale=0.06]{orcid.pdf}\hspace{1mm}Diyin Tang}\\
        Beihang University\\
	\texttt{tangdiyin@buaa.edu.cn}\\
}

\renewcommand{\headeright}{Technical Report}
\renewcommand{\shorttitle}{\textit Coverage-Guaranteed Speech Emotion Recognition}

\hypersetup{
pdftitle={Coverage-Guaranteed Speech Emotion Recognition via Calibrated Uncertainty-Adaptive Prediction Sets},
pdfsubject={q-bio.NC, q-bio.QM},
pdfauthor={David S.~Hippocampus, Elias D.~Striatum},
pdfkeywords={First keyword, Second keyword, More},
}

\begin{document}
\maketitle

\begin{abstract}
Road rage, often triggered by emotional suppression and sudden outbursts, significantly threatens road safety by causing collisions and aggressive behavior. Speech emotion recognition technologies can mitigate this risk by identifying negative emotions early and issuing timely alerts. However, current SER methods, such as those based on hidden markov models and Long short-term memory networks, primarily handle one-dimensional signals, frequently experience overfitting, and lack calibration, limiting their safety-critical effectiveness.
We propose a novel risk-controlled prediction framework providing statistically rigorous guarantees on prediction accuracy. This approach employs a calibration set to define a binary loss function indicating whether the true label is included in the prediction set. Using a data-driven threshold $\beta$, we optimize a joint loss function to maintain an expected test loss bounded by a user-specified risk level $\alpha$.
Evaluations across six baseline models and two benchmark datasets demonstrate our framework consistently achieves a minimum coverage of $1 - \alpha$, effectively controlling marginal error rates despite varying calibration-test split ratios (e.g., 0.1). The robustness and generalizability of the framework are further validated through an extension to small-batch online calibration under a local exchangeability assumption. We construct a non-negative test martingale to maintain prediction validity even in dynamic and non-exchangeable environments. Cross-dataset tests confirm our method's ability to uphold reliable statistical guarantees in realistic, evolving data scenarios.
\end{abstract}


\section{Introduction}
In recent years, road rage has emerged as a growing global traffic safety concern. With rapid urbanization and increasing vehicle ownership, heightened traffic congestion and driving stress have led to a rise in extreme emotional responses among drivers. 
Empirical studies indicate that road rage not only contributes significantly to traffic accidents but also intensifies conflicts between drivers and other road users, thereby posing a serious threat to public safety. According to the National Highway Traffic Safety Administration (NHTSA)~\cite{vachal2024north,world2023global}, over one-third of traffic accidents are associated with emotional loss of control by drivers—a proportion that continues to rise.

Speech Emotion Recognition technology~\cite{wani2021comprehensive,de2023ongoing}, which identifies emotional states through the analysis of vocal signals, has increasingly been recognized as an effective approach to mitigating road rage and other emotion-induced driving risks.~\cite{liu2024primary} 
By capturing vocal characteristics in real time and inferring the driver's emotional state, SER enables timely feedback to intelligent traffic management systems, facilitating the early detection and intervention of potentially hazardous driving behaviors. 
Beyond its applications in smart vehicles for precise emotion monitoring, SER has also shown promise in broader emotion-centric domains such as customer service and mental health assessment~\cite{song2023daily,li2024generating}. 
Given its cross-disciplinary relevance, SER is rapidly gaining traction as a core research focus in intelligent transportation systems, autonomous driving technologies, and affective computing.

With the rapid advancement of deep learning, speech emotion recognition based on convolutional neural networks (CNNs) has achieved notable success. In this study, we utilize Mel-spectrogram features to represent speech signals and differentiate among emotional categories through comparative experiments across models.
Despite these advancements, CNN-based SER models remain susceptible to overfitting during training, leading to degraded generalization on previously unseen data.~\cite{du2025speech,zhang2025sparse} Moreover, most deep learning models lack inherent mechanisms to quantify predictive uncertainty, making their outputs potentially unreliable in real-world applications.
\begin{figure}[t] 
  \centering 
  \includegraphics[width=1.0\textwidth]{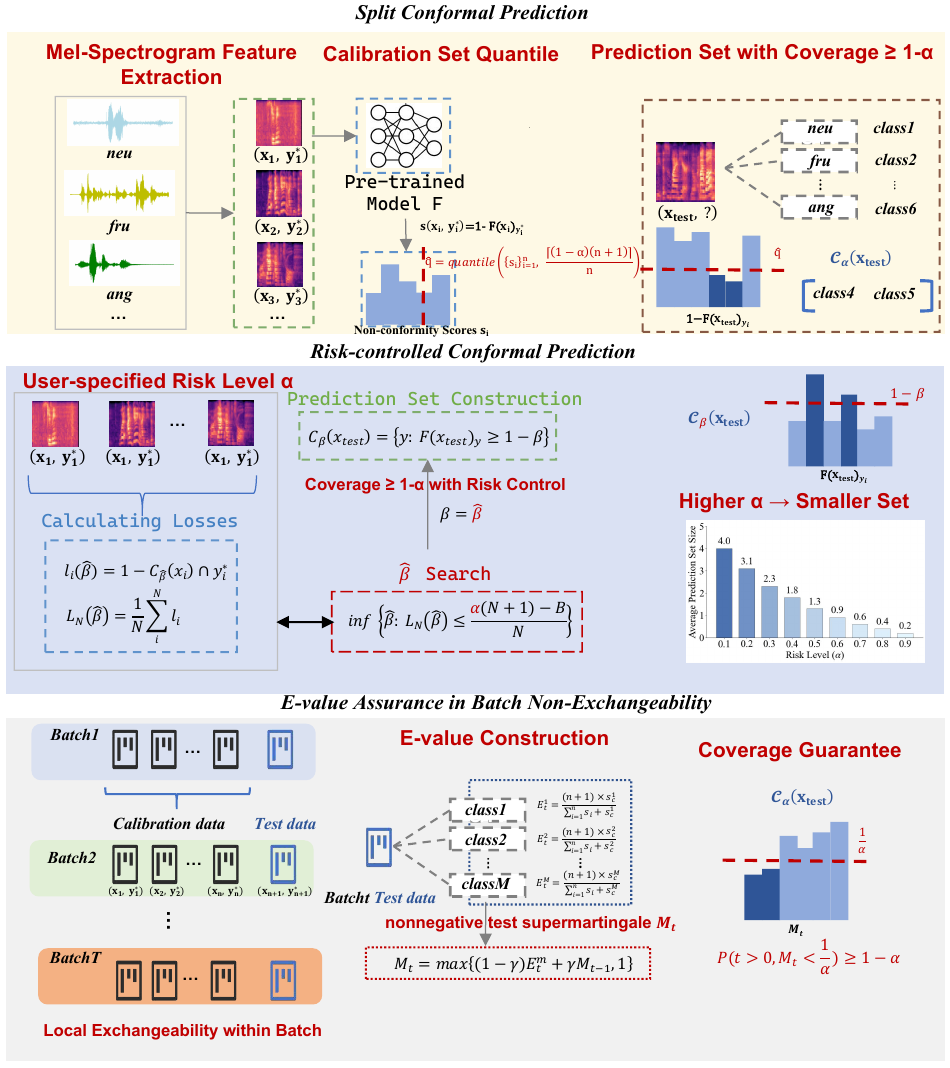} 
  \caption{Overall framework.} 
  \label{fig:verall framework} 
\end{figure}
Recent research has highlighted the issue of miscalibration: Although CNNs often achieve high classification accuracy in SER tasks, their prediction confidence tends to be overestimated. For instance, Zhao et al.~\cite{zhao2023speech} demonstrated that this overconfidence substantially increases the risk of misclassification, particularly for low-confidence samples.
Thus, improving the uncertainty quantification of SER models has become essential to enhancing their reliability and robustness~\cite{psaros2023uncertainty}.

To address these challenges, Conformal Prediction (CP) provides a statistically rigorous framework for uncertainty quantification and prediction calibration~\cite{vovk2005algorithmic}. CP constructs prediction sets that are guaranteed to contain the true label with a user-defined confidence level, thus enhancing model trustworthiness in safety-critical applications.
Our overall framework is illustrated in Figure~\ref{fig:verall framework}. In the context of SER~\cite{li2024meltrans}, we extract Mel-spectrogram features and train five neural network architectures with comparable parameter sizes on the IEMOCAP dataset as base models.
Following model training, the CP procedure is applied in three steps: (1) defining nonconformity scores for each prediction; (2) computing a calibration quantile $\hat{q}$ from a held-out calibration set based on a user-specified risk level $\alpha$; and (3) generating prediction sets for test samples that satisfy the marginal coverage guarantee.
By outputting emotion label sets with formal confidence guarantees, our approach establishes a principled statistical foundation for improving the reliability of speech emotion recognition systems.

While conformal prediction provides formal coverage guarantees for prediction sets, it primarily focuses on marginal coverage and does not account for task-specific performance criteria.
To address this limitation, the risk-controlled conformal prediction (RCCP) framework~\cite{angelopoulos2022conformal} extends traditional CP by bounding the expected value of arbitrary monotonic loss functions, thereby enabling principled risk management in complex real-world tasks.
This generalization maintains statistical rigor while offering flexibility in defining application-specific loss metrics, making the approach broadly applicable across diverse domains.
By combining CP with risk control, the framework preserves coverage guarantees while aligning predictions with user-defined performance requirements. 
However, conventional CP methods assume exchangeability among data points—a condition that holds in controlled environments but is frequently violated in real-world scenarios, such as speech emotion recognition, where data distributions often shift across speakers, sessions, or acoustic conditions.
In such non-exchangeable settings, standard CP methods may fail to uphold their theoretical guarantees, thus compromising reliability. 
To address this challenge, we propose a novel framework that extends CP to dynamic, non-exchangeable environments via a mini-batch online calibration mechanism.
Specifically, we relax the global exchangeability assumption by introducing a local exchangeability assumption within individual mini-batches.
By constructing a non-negative test supermartingale, our method adaptively calibrates prediction sets in an online manner, ensuring valid coverage even under distributional shifts between batches—such as variations in speaker identity or environmental noise.
This framework provides a statistically sound and practically robust solution for deploying CP-based uncertainty quantification in dynamic real-world SER applications.

To validate the effectiveness of our framework, we conducted a series of empirical studies. Preliminary evaluations on the TESS dataset compared multiple feature extraction methods, with Mel-spectrograms demonstrating superior performance and thus selected for downstream tasks. 
We then trained several neural network architectures (e.g., ResNet, MobileNetV3, ShuffleNetV2) on IEMOCAP, where a clear overfitting problem emerged, despite perfect training accuracy, test performance degraded significantly.
To ensure prediction reliability, we integrated the CP framework, conducting both within-dataset and cross-dataset experiments. Results confirmed that CP consistently maintained desired coverage levels, even under data distribution shifts, highlighting its robustness. 
Moreover, we observed that prediction set size serves as an effective indicator of model uncertainty~\cite{wang2025word}, prompting the use of a risk-controlled loss function to adaptively regulate prediction confidence.
Finally, to address non-exchangeable settings, we introduced a mini-batch martingale-based calibration strategy. Experiments with mismatched calibration and test datasets (IEMOCAP to TESS) demonstrated that, unlike standard CP, our approach preserved high coverage under local exchangeability assumptions, confirming its practical utility in dynamic environments.

\section{Related Work}
\subsection{Speech Recognition}
The evolution of speech recognition technology has progressed from early template matching systems in the 1950s to modern deep learning approaches. Traditional models, such as hidden Markov models and Gaussian mixture models (GMMs), enabled statistical modeling of speech sequences but struggled with noise and variability~\cite{rabiner1989tutorial,deng2014deep}.
The advent of deep learning transformed the field: convolutional and recurrent neural networks, including long short-term memory architectures, enabled end-to-end training and improved robustness in real-world scenarios~\cite{hinton2012deep,graves2013hybrid}.
In the context of speech emotion recognition, deep learning has similarly replaced hand-crafted feature engineering with automatic representation learning.~\cite{uddin2020emotion,sun2024caterpillar} As reviewed by Schuller~\cite{schuller2018speech}, hybrid CNN-LSTM models now form the backbone of state-of-the-art SER systems, capable of capturing complex emotional nuances in speech data.
Today, deep learning stands as the foundation of both speech recognition and emotion understanding, powering a wide range of applications from intelligent assistants to mental health monitoring.

\subsection{Conformal Prediction}
Conformal Prediction ~\cite{angelopoulos2021gentle,ye2024benchmarking,barber2023conformal,wang-etal-2024-conu,sun2023copula,wang2025sample} is an uncertainty quantification framework that offers finite-sample statistical guarantees for predictive models.
Unlike traditional point predictions, CP outputs a prediction set—a subset of possible labels—that is guaranteed to contain the true label with a user-specified confidence level (e.g., 90\%). 
The core idea of CP is to compute a nonconformity score that quantifies how unusual a predicted label is, given the model’s behavior. For classification tasks, this score is often based on the model's confidence in the correct class (e.g., 1 minus the predicted probability of the true label); for regression, it typically corresponds to the absolute error between predicted and true values. 
During calibration, CP uses a held-out calibration set to estimate the distribution of nonconformity scores. A quantile threshold is then determined based on the desired coverage level. At test time, for each new input, the model evaluates nonconformity scores across candidate labels, and the prediction set includes all labels with scores below the calibrated threshold. 
Importantly, CP makes no assumptions about the underlying data distribution, requiring only that the calibration and test samples are exchangeable. This distribution-free property enables CP to provide reliable coverage guarantees across a wide range of models and applications.

\section{Method}
\begin{figure}[t] 
  \centering 
  \includegraphics[width=1\textwidth]{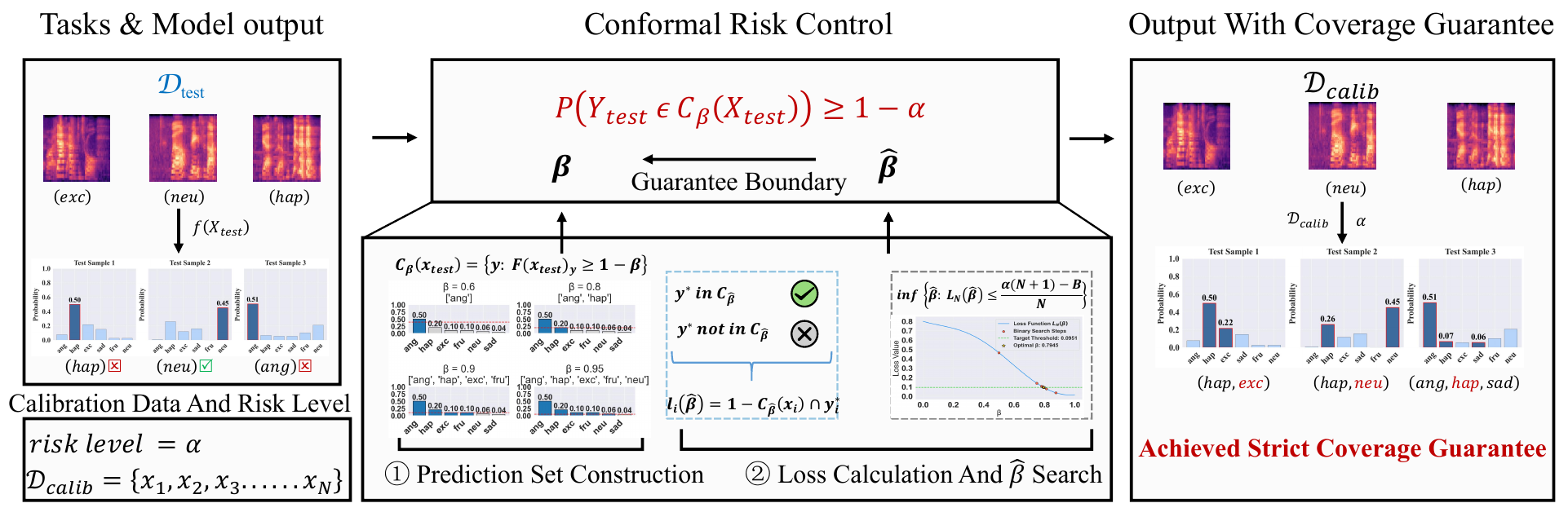} 
  \caption{Illustration of the Risk-controlled Conformal Prediction Framework} 
  \label{fig:example} 
\end{figure}
\subsection{Preliminaries}

As a statistical consistency prediction method, the Split Conformal Prediction(SCP) framework shows unique advantages in the reliability evaluation of machine learning models by converting heuristic uncertainty measures into prediction sets with strict probability guarantees. Its core idea is based on the assumption of data exchangeability and combines the quantile calibration of nonconformity scores to achieve coverage probability control of unknown samples. The theoretical basis and algorithm flow of SCP will be systematically explained below.

\subsubsection{Theoretical Foundations and Algorithm Process}
\paragraph{1. Problem Definition and Input Conditions}

\begin{itemize}
    \item \textbf{Input Data}:
    \begin{itemize}
        \item Calibration dataset $\{(X_t, Y_t^*)\}_{t=1}^n$, satisfying the independent and identically distributed (i.i.d.) assumption.
        \item Pretrained classification model $\hat{f}(\cdot)$, with output as a class probability vector $\hat{f}(X_t) \in [0,1]^K$, where the true label probability is $\hat{f}(X_t)_{Y_t^*}$.
    \end{itemize}
    \item \textbf{Objective Function}: Construct the prediction set $\mathcal{C}(X_{\text{test}})$ such that the coverage probability of the true label satisfies:
    \begin{equation}
    \mathbb{P}(Y_{\text{test}}^* \in \mathcal{C}(X_{\text{test}})) \geq 1 - \alpha
    \end{equation}
\end{itemize}

\paragraph{2. Nonconformity Score Definition and Quantile Calibration}
The nonconformity score $s_t$ reflects the model's uncertainty in predicting the true label, defined as:
\begin{equation}
s_t = 1 - \hat{f}(X_t)_{Y_t^*}
\end{equation}
Sort the scores of the calibration set in ascending order: $\{s_1 \leq s_2 \leq \cdots \leq s_n\}$, and Get the $\frac{\lceil (n+1)(1-\alpha) \rceil}{n} 
$ quantile from $\{s_t\}_{t=1}^n$:
\begin{equation}
\hat{q} = \inf \left\{ q : \frac{|\{t : s_t \leq q\}|}{n} \geq \frac{(n+1)(1-\alpha)}{n} \right\} = s_{\frac{\lceil (n+1)(1-\alpha) \rceil}{n}}
\end{equation}

\textbf{Key Derivation}: Based on the exchangeability assumption, the nonconformity score $s_{\text{test}}$ of the test sample is uniformly distributed among the calibration scores in the sorted sequence. The coverage probability strictly satisfies:
\begin{equation}
\mathbb{P}(s_{\text{test}} \leq \hat{q}) = \frac{\lceil (n+1)(1-\alpha) \rceil}{n+1} \geq 1 - \alpha
\end{equation}
Here, the ceiling function $\lceil \cdot \rceil$ ensures the conservativeness of the threshold, thus guaranteeing the lower bound of the coverage probability.

\paragraph{3. Prediction Set Construction and Coverage Validation}

For the test sample $X_{\text{test}}$, the prediction set includes all classes satisfying the following condition:
\begin{equation}
\mathcal{C}(X_{\text{test}}) = \left\{ y \in [K] : 1 - \hat{f}(X_{\text{test}})_{y} \leq \hat{q} \right\}
\end{equation}

\textbf{Example Operation}:
\begin{itemize}
    \item If the predicted probability for a class $\hat{f}(X_{\text{test}})_{y} \geq 1 - \hat{q}$, the class is included in the prediction set.
    \item The final prediction set size is dynamically adjusted, balancing confidence and classification granularity.
\end{itemize}

\subsubsection{Mathematical Proof Supplement}

\textbf{Theorem (Marginal Coverage Guarantee)}: Under the data exchangeability assumption, the SCP framework satisfies:
\begin{equation}
\mathbb{P}(Y_{\text{test}}^* \in \mathcal{C}(X_{\text{test}})) \geq 1 - \alpha
\end{equation}

\textbf{Proof}: Let the combined sequence of calibration and test scores be $s_{(1)} \leq s_{(2)} \leq \cdots \leq s_{(n+1)}$, with the rank $k$ of the test score $s_{\text{test}}$ following a uniform distribution $k \sim \text{Uniform} \{1, \dots, n+1\}$. Select the threshold $\hat{q} = s_{(k^*)}$, where $k^* = \lceil (n+1)(1-\alpha) \rceil$, then:
\begin{equation}
\mathbb{P}(s_{\text{test}} \leq \hat{q}) = \mathbb{P}(k \leq k^*) = \frac{k^*}{n+1} \geq 1 - \alpha
\end{equation}

\subsection{Risk-controlled Conformal Prediction in speech emotion recognition task}
To adapt the framework of split conformal prediction to the
speech emotion recognition task for user - specified guarantees
of task - specific performance,we have developed a risk-controlled conformal prediction framework, as shown in the figure \ref{fig:example}, which explains how the framework works, we define a loss function for each calibration data, formulated as
\begin{equation}
   \ell(C(X_t), Y_t) = \mathbf {1}\{Y_t \notin C(X_t)\}
\end{equation}
RCCP provides statistical guarantees for high - risk classification tasks through the exchangeable data assumption and custom - defined loss functions. Given \( N \) calibration data points and 1 test data point (with a total of \( K \) classes), RCCP first constructs a prediction set for the sample \( x_i \):  
\begin{equation}
C_{\beta}(x_i)=\left\{y:F(x_i)_y\geq 1 - \beta\right\}
\end{equation}  
where \( F(x_i) \) is the probability distribution output by the model, and \( \beta \) controls the confidence threshold. The task - specific loss is defined as \( l_i = 1-{|C_{\beta}(x_i)\cap y_i^*|} \) (if the true label \( y_i^*\notin C_{\beta}(x_i) \), then \( l_i = 1 \)), and its expectation must satisfy \( \mathbb{E}[l_{\text{test}}]\leq\alpha \). Based on data exchangeability, the expectation of the test loss can be expressed as:  
\begin{equation}
\mathbb{E}[l_{\text{test}}(\beta)]=\frac{N L_N(\beta)+l_{\text{test}}(\beta)}{N + 1}\leq\alpha
\end{equation}
where \( L_N(\beta)=\frac{1}{N}\sum_{i = 1}^N l_i \) is the average loss of the calibration set. By solving for the optimal threshold  
\begin{equation}
\hat{\beta}=\inf\left\{\beta:\frac{N L_N(\beta)+B}{N + 1}\leq\alpha\right\}
\end{equation}
(where \( B \) is the task - related loss upper bound, e.g., \( B = 1 \)), \( \hat{\beta} \) is determined and applied to the test data to construct \( C_{\hat{\beta}}(x_{\text{test}}) \), ensuring risk control. This framework flexibly designs loss functions (such as coverage loss \( l_i=1 - \) coverage or sparsity loss \( l_i = |C_{\beta}(x_i)| \)), transforming abstract task metrics (like false positive rate and prediction set size) into an optimizable form, thereby achieving task - specific metric guarantee.

\begin{algorithm}[t]
\caption{Conformal Prediction with Nonnegative Test Supermartingale}\label{alg:smcs}
\begin{algorithmic}[1]
\Require 
    \Statex Pre-trained classifier $f: \mathcal{X} \to \Delta^{K-1}$
    \Statex Significance level $\alpha \in (0,1)$
    \Statex Mixing rate $\gamma \in (0,1)$
\State \textbf{Initialize} 
    \Statex Supermartingale $M_0 \gets 1$ 
\Procedure{OnlinePrediction}{Test stream $\{(x_t,y_t)\}_{t=1}^T$, $\mathcal{D}_{\text{cal}}$}
\For{$t=1$ \textbf{to} $T$}
    \State Sample $\mathcal{B}_t \subset \mathcal{D}_{\text{cal}}$ with $|\mathcal{B}_t|=5$ 
    \State $\bm{p}_t \gets f(x_t)$ \Comment{$\bm{p}_t \in \Delta^{K-1}$}
    \State Compute $\{s_i\}_{i=1}^n$ where $s_i = 1 - \bm{p}_{z_i}(y_i)\ \forall z_i \in \mathcal{B}_t$
    
    \State $C_t \gets \emptyset$
    \For{$m=1$ \textbf{to} $K$}
        \State $s^m_t \gets 1 - \bm{p}_t(m)$
        \State $E^m_t \gets \frac{(n+1)s^m_t}{\sum_{s_i \in \mathcal{B}_t} s_i + s^m_t}$
        \State $M^m_t \gets \max\big\{(1-\gamma)E^m_t + \gamma M_{t-1},\ 1\big\}$
        \If{$M^m_t < 1/\alpha$}
            \State $C_t \gets C_t \cup \{m\}$
        \EndIf
    \EndFor
    
    \State $E^{y_t}_t \gets \frac{(n+1)(1 - \bm{p}_t(y_t))}{\sum_{s_i \in \mathcal{B}_t} s_i + (1 - \bm{p}_t(y_t))}$
    \State $M_t \gets \max\big\{(1-\gamma)E^{y_t}_t + \gamma M_{t-1},\ 1\big\}$
\EndFor
\EndProcedure
\end{algorithmic}
\end{algorithm}

\subsection{Comparison of two frameworks} 
RCCP  has a core drawback compared to SCP  in terms of lower computational efficiency, mainly due to its need for multiple traversals of the calibration data to optimize the threshold. Specifically: 
The threshold calculation of SCP only requires a single sorting of the non-conformity scores \( s_t = 1 - \hat{f}(X_t)_{Y^*_t} \) on the calibration set, and taking the quantile \( \hat{q} \), with a time complexity of \( O(N \log N) \). 
RCCP needs to traverse candidate thresholds \( \beta \). For each candidate value, it calculates the average loss on the calibration set \( L_N(\beta) = \frac{1}{N} \sum_{i=1}^{N} l_i \), and then verifies the condition \( \frac{N L_N(\beta) + B}{N + 1} \leq \alpha \). Assuming the number of candidate \( \beta \) is \( M \), its time complexity is \( O(M \cdot N) \)
Even with binary search optimization (iteration times \( O(\log M) \)), it still requires multiple complete traversals of the calibration set, leading to a significant increase in computational load. 
This efficiency bottleneck is particularly prominent in large-scale calibration sets (\( N \gg 1 \)) or high-real-time scenarios (such as autonomous driving), while the lightweight single sorting of SCP has more advantages. The flexibility of RCCP comes at the cost of efficiency, and one needs to balance the choice according to task requirements.

\subsection{Mini-batch conformal prediction theory under Non-exchangeable Data} 
In light of the limitations of traditional conformal prediction, which relies on global data exchangeability~\cite{barber2023conformal,KangGYS024,wang2025sconu}, we propose the assumption of within - batch local exchangeability. We prove that even when there are differences in data distributions between batches (i.e., batch - to - batch exchangeability may not hold), we can still strictly guarantee the global coverage probability by constructing a non - negative test supermartingale with a forcing operation. The framework is shown in the algorithm \ref{alg:smcs}. This framework breaks through the strong dependence of classical theories on global data exchangeability, providing universal theoretical support for small - batch scenarios such as time - series data and online learning.

E - value~\cite{gauthier2025values,Subhabrata2022evalue}, as a non - negative random variable satisfying \(E[E]\leq1\), can directly construct conformal sets through probability inequalities, offering a more flexible uncertainty quantification tool. For the calibration set \(D_{cal}=\{(X_{i},Y_{i})\}_{i = 1}^{n}\) and the test sample \((X_{n + 1},Y_{n+1})\), the E - value is defined as: 
\begin{equation}
    E=\frac{s_{n+1}}{\frac{1}{n + 1}\sum_{i=1}^{n+1}s_{i}}
\end{equation}
where \(s_{i}=s(X_{i},Y_{i})\) and \(s\) is a non - conformity score (such as cross - entropy loss) with non - negative values. Under the assumption of within - batch data exchangeability, the conditional expectation of the E - value satisfies \(E[E_{t}|F_{t - 1}]=1\), which provides a crucial basis for constructing the martingale process.

The filtration \(\{F_{t}\}_{t\geq0}\) is defined as the \(\sigma\) - algebra generated by all historical data up to the \(t\) batch. Specifically, we define the filtration
\begin{equation}
F_{t}=\sigma(s_{1}^{1},\cdots,s_{n_{1}}^{1},s_{n_{1}+1}^{1},\cdots,s_{1}^{t},\cdots,s_{n_{t}}^{t},s_{n_{t}+1}^{t})
\end{equation}

For all \(t\geq0\), the sequence of random variables \(\{M_{t}\}_{t\geq0}\) is defined as:
\begin{equation}
M_{t}=\max\{(1 - \gamma)E_{t}+\gamma M_{t - 1},1\}
\end{equation}
where
\begin{equation}
\small
E_{t}=\frac{(n_{t}+1)\times s_{n_{t}+1}}{\sum_{i = 1}^{n_{t}}s_{i}+s_{n_{t}+1}}
\end{equation}
We will prove that \(\{M_{t}\}_{t\geq0}\) is a non - negative test supermartingale.
\begin{itemize}
    \item \textbf{Initial condition}: First, we define \(M_{0}=1\). Since the non - conformity scores we calculate are positive, it is obvious that for all \(t\geq0\), \(M_{t}\geq0\). The remaining task is to prove that \(\{M_{t}\}_{t\geq0}\) is a supermartingale. A non - negative test supermartingale needs to satisfy: \(M_{0}=1\) (already satisfied), \(M_{t}\geq0\) and is \(F_{t}\) - measurable, and \(E[M_{t}|F_{t - 1}]\leq M_{t - 1}\).
    \item \textbf{Measurability}: To determine whether \(M_{t}\) is measurable after defining the filtration, we use mathematical induction. When \(t = 0\), \(M_{0}=1\). Since a constant function is measurable with respect to any \(\sigma\) - algebra, \(M_{0}\) is \(F_{0}\) - measurable, where \(F_{0}=\{\varnothing,\Omega\}\) is the trivial \(\sigma\) - algebra. Assume that \(M_{t - 1}\) is \(F_{t - 1}\) - measurable. Because \(F_{t - 1}\subseteq F_{t}\), \(M_{t - 1}\) is also \(F_{t}\) - measurable. For the \(t\) batch, \(E_{t}\) only depends on the non - conformity scores \(s_{1}^{t},\cdots,s_{n_{t}}^{t},s_{n_{t}+1}^{t}\) of the \(t\) - th batch, and these scores are all in \(F_{t}\), so \(E_{t}^{m_{-}true}\) is \(F_{t}\) - measurable. According to the update rule, let \(Z_{t}=(1 - \gamma)E_{t}+\gamma M_{t - 1}\). Since \(E_{t}\) and \(M_{t - 1}\) are both \(F_{t}\) - measurable, the linear combination \(Z_{t}\) is also \(F_{t}\) - measurable. Let \(A_{t}=\{\omega\in\Omega:Z_{t}(\omega)<1\}\). Since \(Z_{t}\) is \(F_{t}\) - measurable, \(A_{t}\in F_{t}\). We can rewrite \(M_{t}\) as \(M_{t}=1\cdot I_{A_{t}}+Z_{t}\cdot I_{A_{t}^{c}}\), where \(I_{A_{t}}\) is the indicator function of the set \(A_{t}\). The indicator functions \(I_{A_{t}}\) and \(I_{A_{t}^{c}}\) are both \(F_{t}\) - measurable, and the products of the constant 1 and the measurable function \(Z_{t}\) with the indicator functions are also \(F_{t}\) - measurable. Thus, their sum \(M_{t}\) is \(F_{t}\) - measurable. In conclusion, by mathematical induction, we have proved that for the defined filtration \(\{F_{t}\}_{t\geq0}\), \(M_{t}\) is \(F_{t}\) - measurable.
    \item \textbf{Decreasing conditional expectation}: Under the assumption of within - batch data exchangeability, \(E[E_{t}|F_{t - 1}]=1\). Then the conditional expectation of the supermartingale is 
    \begin{equation}
    E[M_{t}|F_{t - 1}]=(1-\lambda)E[E_{t}|F_{t - 1}]+\lambda M_{t - 1}=(1-\lambda)+\lambda M_{t - 1}
    \end{equation}
    Since \(M_{t - 1}\geq1\) and \(\lambda\in(0,1)\), we have \(E[M_{t}|F_{t - 1}]\leq M_{t - 1}\), which satisfies the property of a supermartingale.
\end{itemize}
Ville's inequality states that for a non - negative supermartingale \(M_{t}\), \(P(\forall t\geq0,M_{t}<\frac{1}{\alpha})\geq1-\alpha\). The construction rule of the prediction set is to include the corresponding category in the prediction set when \(M_{t}<\frac{1}{\alpha}\). This means that the probability that the true label falls into the prediction set is at least \(1-\alpha\), thus theoretically guaranteeing the coverage rate of the prediction set. Let \(x_{t}\) be the \(t\) - th test sample and \(Y\) be the set of all possible categories. For the test sample \(x_{t}\), its prediction set \(\hat{C}(x_{t})\) can be expressed as:
\begin{equation}
\small
\hat{C}(x_{t})=\left\{m\in Y:\max\{(1 - \gamma)\frac{(n_{t}+1)\cdot s(x_{t},m)}{\sum_{i = 1}^{n_{t}}s_{i}+s(x_{t},m)}+\gamma M_{t - 1},1\}<\frac{1}{\alpha}\right\}
\end{equation}

\section{Experiments}
\subsection{Experimental Settings}
\begin{figure}[t]
    \centering
    \begin{subfigure}[b]{0.9\textwidth} 
        \centering
        \includegraphics[width=1\textwidth]{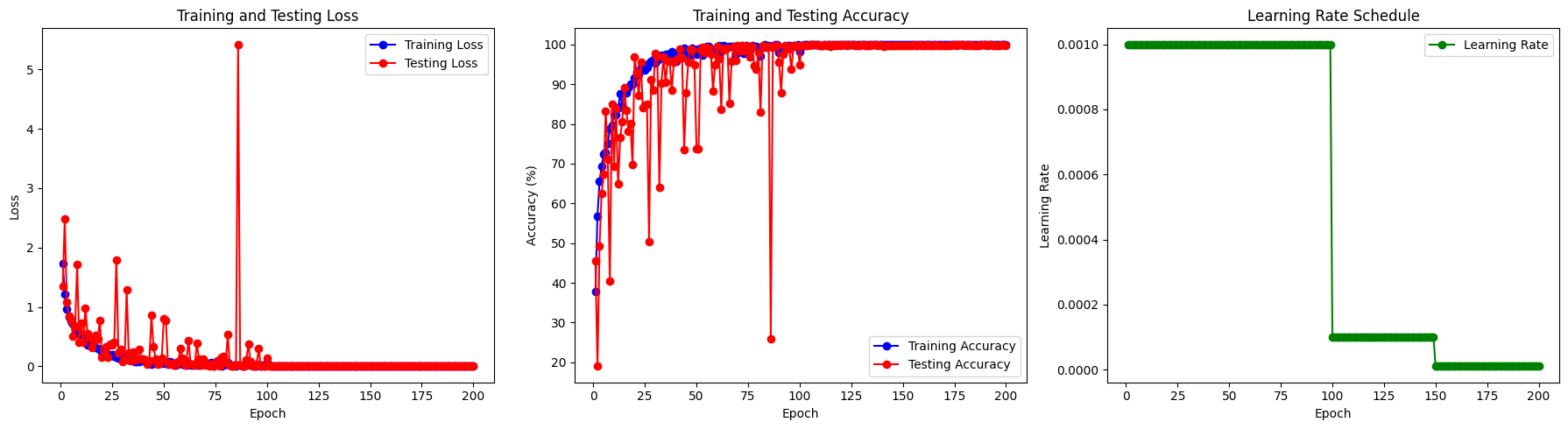} 
        \caption{Training and testing results of Mel-spectrogram features in basic CNN model}
        \label{fig:trainsub1}
    \end{subfigure}
    
    \vspace{0.1cm} 
    
    \begin{subfigure}[b]{0.9\textwidth}
        \centering
        \includegraphics[width=1\textwidth]{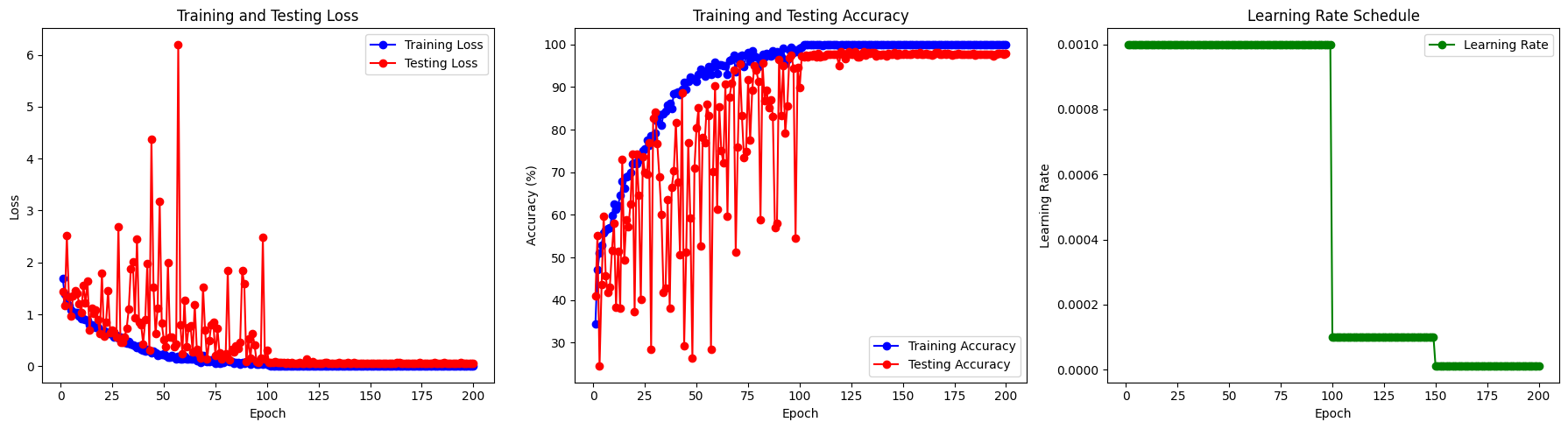} 
        \caption{Training and testing results of chromaticity features in the basic CNN model}
        \label{fig:trainsub2}
    \end{subfigure}
    
    \vspace{0.1cm}
    
    \begin{subfigure}[b]{0.9\textwidth}
        \centering
        \includegraphics[width=1\textwidth]{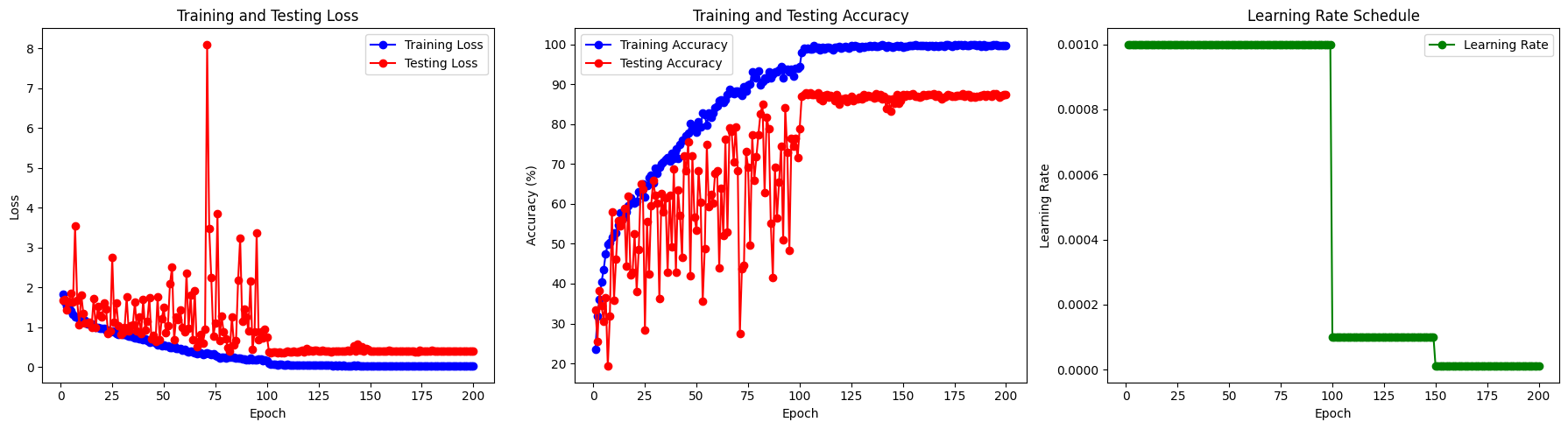} 
        \caption{Training and testing results of contrast features in the basic CNN model}
        \label{fig:trainsub3}
    \end{subfigure}
    
    \caption{Comparative analysis of the performance of different speech features in basic CNN models based on the TESS dataset}
    \label{fig:total}
\end{figure}

\subsubsection{Datasets}
This study conducts model training and evaluation based on two multimodal emotion datasets, IEMOCAP  and TESS. The IEMOCAP dataset contains 12 hours of audio-visual data collected across 5 recording sessions, where ten professional actors (5 male, 5 female) simulate real-life interaction scenarios.~\cite{busso2008iemocap}For model training, six emotion categories are selected: "ang" (anger), "hap" (happiness), "exc" (excitement), "sad" (sadness), "fru" (frustration), and "neu" (neutral). A stratified sampling approach is employed to ensure each category provides 2,000 balanced training samples.  The TESS dataset comprises 2,800 audio files, in which two female actors express seven discrete emotions through the fixed carrier sentence structure "Say the word\_".~\cite{pichora2020toronto}To establish cross-dataset consistency, only emotion categories overlapping with IEMOCAP are retained. Specifically, "Anger", "Happiness", "Sadness", and "Neutral" in TESS are mapped to the corresponding IEMOCAP labels "ang", "hap", "sad", and "neu", respectively. This label harmonization ensures uniform emotional category definitions during model training and evaluation across datasets.

\subsubsection{Base Models}
In this study, we used five convolutional neural network (CNN) architectures with comparable number of parameters and a base CNN network: ResNet (1,167,971), MobileNetV3 (1,166,910), ShuffleNetV2 (1,173,798), SqueezeNet (1,175,610), and GhostNet (1,177,659). These models employ distinct architectural designs to balance computational efficiency and performance. Residual connections in ResNet address the gradient vanishing issue in deep networks~\cite{he2016deep}; MobileNetV3 and ShuffleNetV2 adopt lightweight frameworks tailored for resource-constrained scenarios~\cite{howard2019searching,ma2018shufflenet};
\begin{figure}[H]
    \centering
    
    \begin{subfigure}[b]{0.45\textwidth}
        \includegraphics[width=\textwidth]{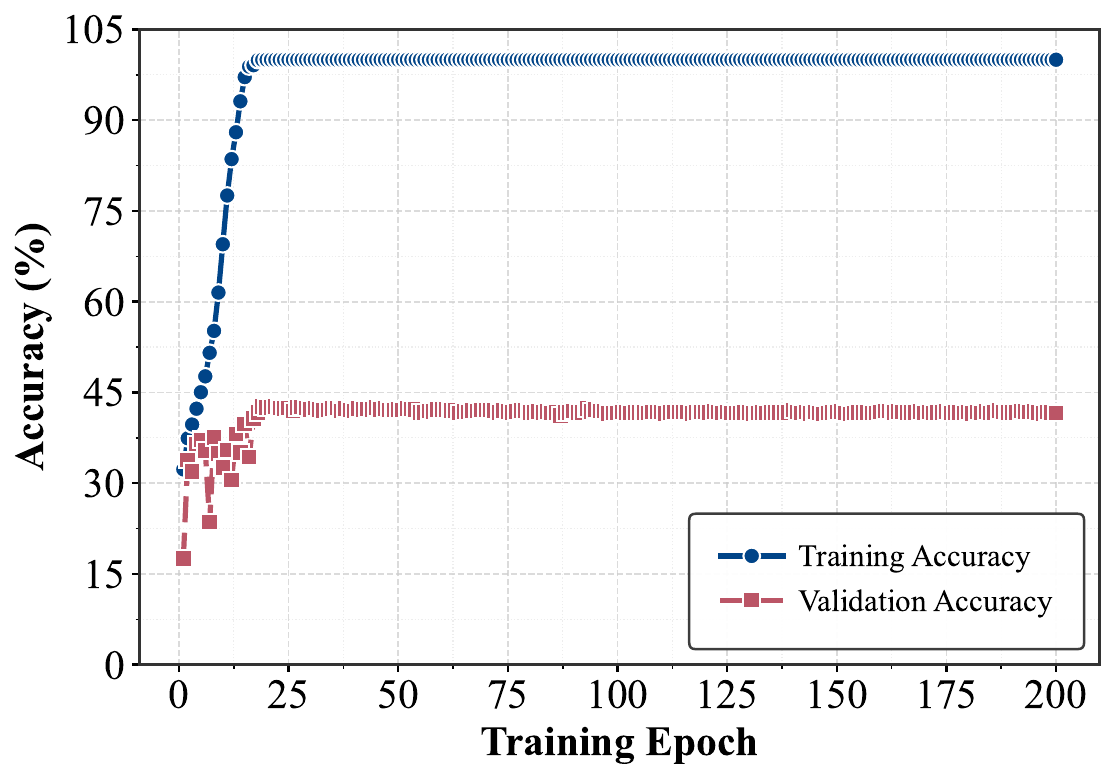}
        \caption{ResNet training and testing Accuracy}
        \label{fig:sub1}
    \end{subfigure}
    \hfill
    \begin{subfigure}[b]{0.45\textwidth}
        \includegraphics[width=\textwidth]{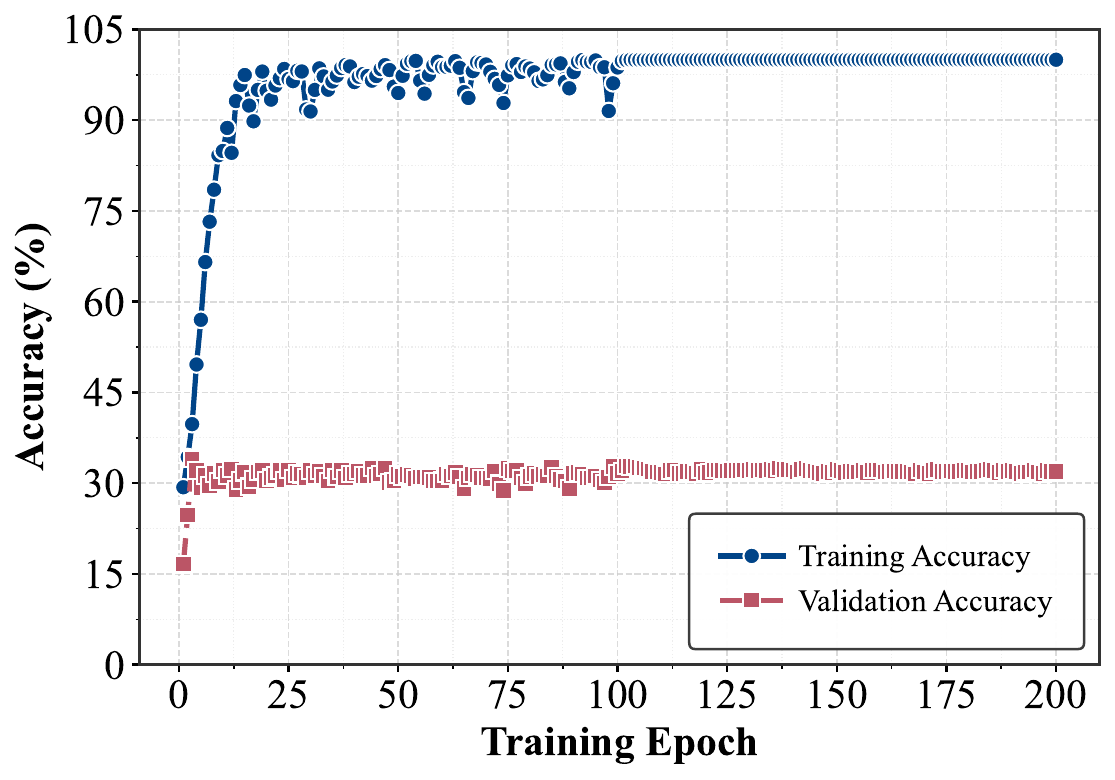}
        \caption{MobileNetV3 training and testing Accuracy}
        \label{fig:sub2}
    \end{subfigure}
    
    \vspace{0.1cm}  
    
    \begin{subfigure}[b]{0.45\textwidth}
        \includegraphics[width=\textwidth]{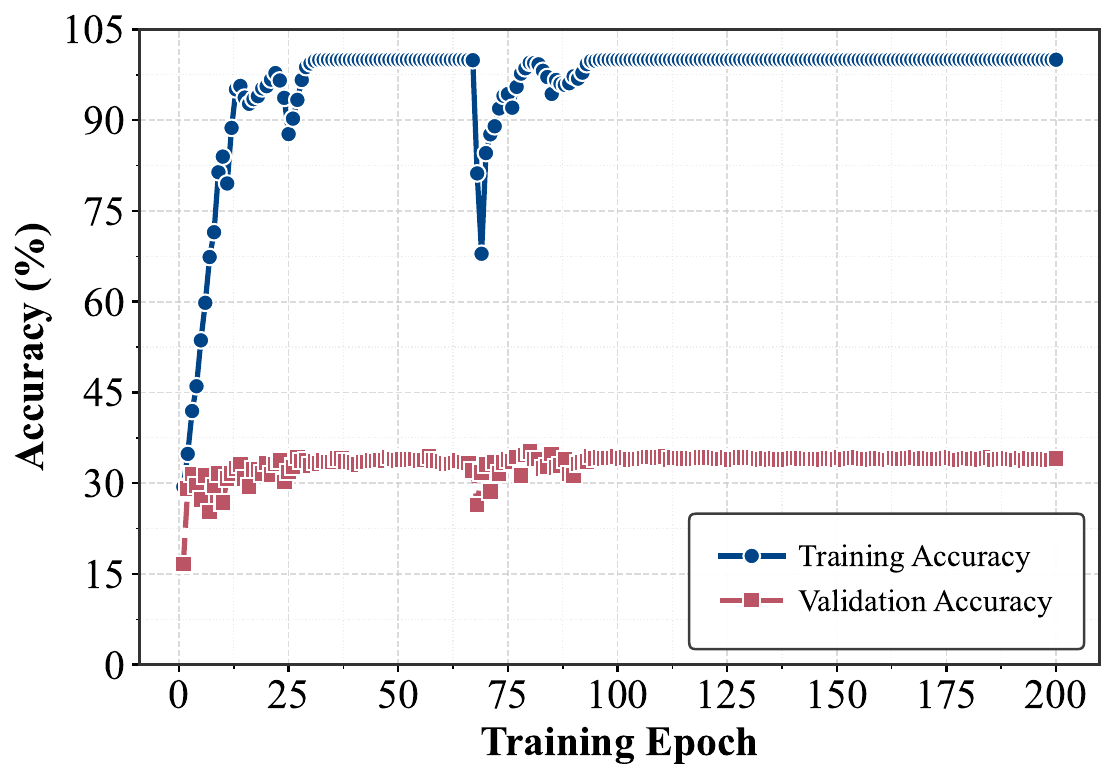}
        \caption{ShuffleNetV2 training and testing Accuracy}
        \label{fig:sub3}
    \end{subfigure}
    \hfill
    \begin{subfigure}[b]{0.45\textwidth}
        \includegraphics[width=\textwidth]{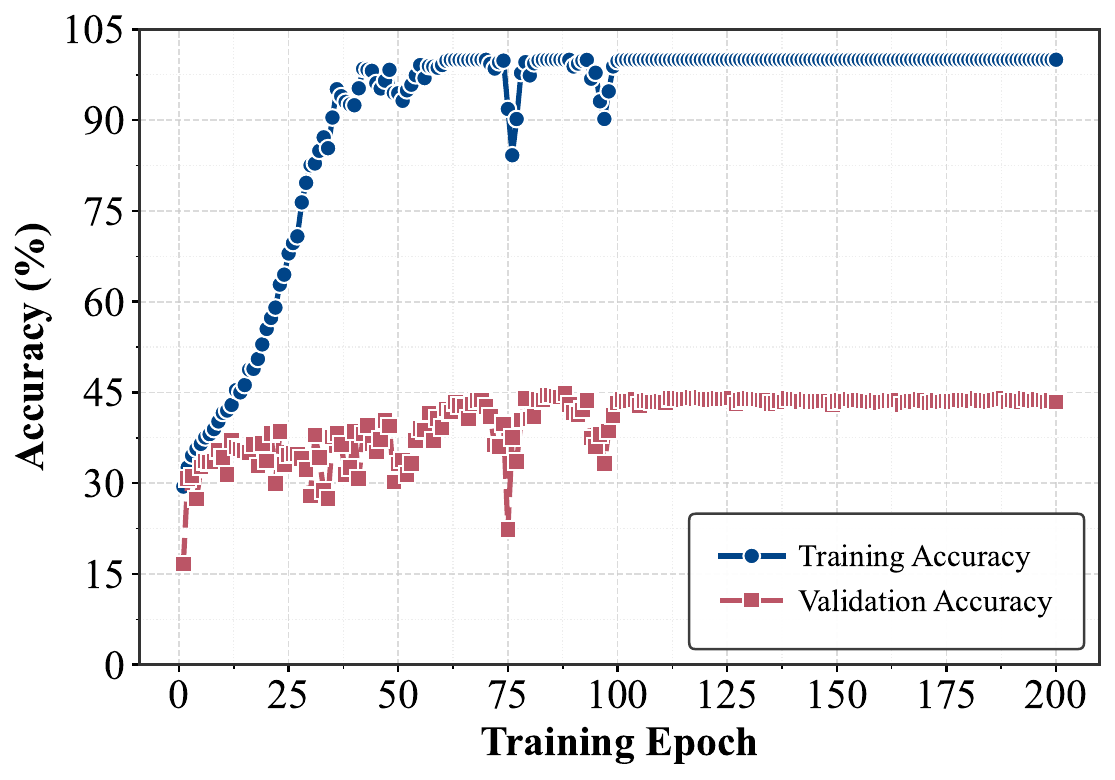}
        \caption{SqueezeNet training and testing Accuracy}
        \label{fig:sub4}
    \end{subfigure}
    
    \vspace{0.1cm}
    
    \begin{subfigure}[b]{0.45\textwidth}
        \centering
        \includegraphics[width=\textwidth]{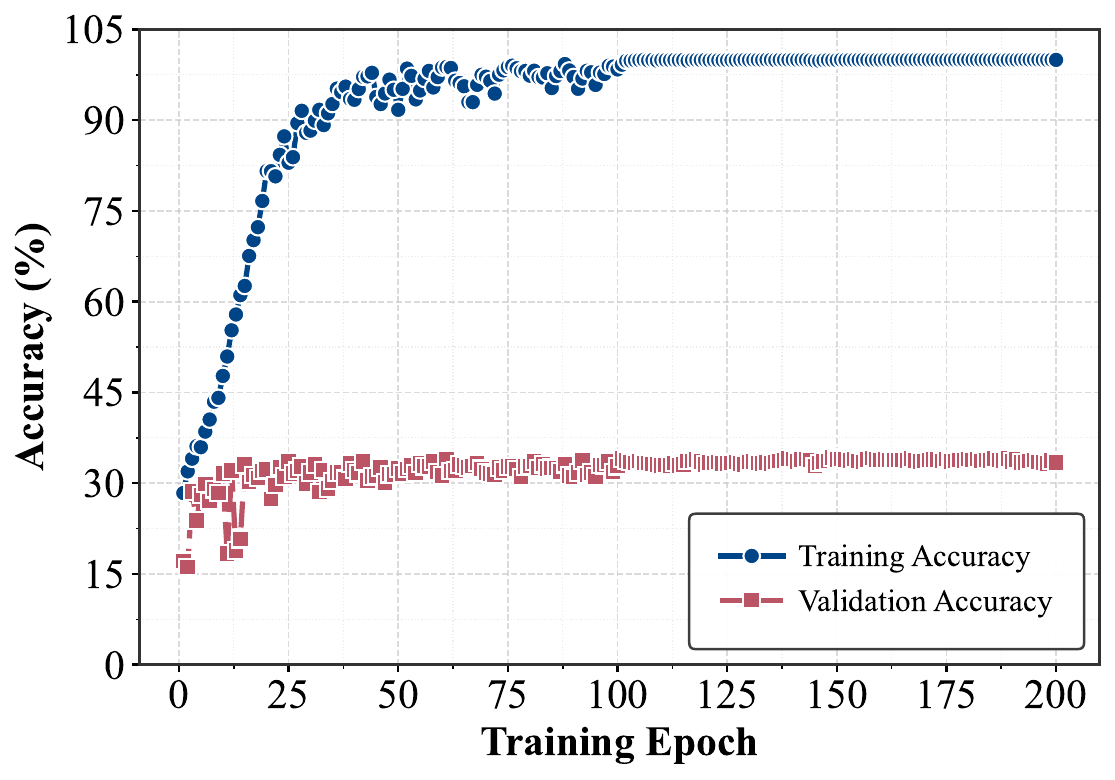}
        \caption{GhostNet training and testing Accuracy}
        \label{fig:sub5}
    \end{subfigure}
    
    \caption{Comparison of training and test set Accuracy of different neural network models}
    \label{fig:group1}
\end{figure}
while SqueezeNet and GhostNet further minimize parameter overhead and computational complexity, making them suitable for embedded and mobile applications~\cite{iandola2016squeezenet,han2020ghostnet}.

\subsubsection{Methods for Speech Feature Extraction}
We compared three speech feature extraction methods: Mel spectrogram, chromaticity feature, and contrast feature. Mel spectrogram is a commonly used feature in speech processing. It maps the spectrum of speech signals to the Mel scale, which simulates the human auditory system's ability to perceive different frequencies. By converting the speech signal into a spectrum obtained by short - time Fourier transform (STFT) and applying Mel filter bank processing, Mel spectrogram can effectively extract important frequency components in speech. Compared with traditional linear spectrum, Mel spectrogram can better capture the acoustic characteristics of speech. Chromaticity feature converts spectrum information into energy distribution of 12 scales, representing the change of pitch. Through these features, the tonality information in speech can be reflected, which is particularly important for emotion recognition because different emotions may affect the pitch and tonality of speech. Contrast feature reflects the energy difference between different frequency bands in the spectrum and can describe the local structure of the spectrum. It helps capture the detailed changes in speech by calculating the energy difference between each frequency band and the energy of adjacent frequency bands. It can reflect the emotional color and emotional intensity changes of speech.

\subsubsection{Evaluation Metrics}
We employ the Empirical Coverage Rate (ECR) to evaluate whether we rigorously control the error rates at various user - specified risk levels. Moreover, we leverage the Average Prediction Set Size  to assess the uncertainty of the model's decision - making and the prediction efficiency of the calibrated prediction sets.

\subsubsection{Hyper - parameters}
We set the split ratio between the calibration set and the test set to 50\% in the IEMOCAP dataset and 50\% in the TESS dataset. We modified the structure of the models to ensure that each model has a similar number of parameters: ResNet (1,167,971 parameters), MobileNetV3 (1,166,910 parameters), ShuffleNetV2 (1,173,798 parameters), SqueezeNet (1,175,610 parameters), and GhostNet (1,177,659 parameters) to evaluate the uncertainty of different models with the same parameter size.
\begin{table}[t]
\centering
\setlength{\tabcolsep}{3pt}
\setlength{\textfloatsep}{4pt} 
\setlength{\abovecaptionskip}{4pt}  
\setlength{\belowcaptionskip}{0pt}
\caption{Classification Accuracy Comparison Across Models on IEMOCAP and TESS Datasets}
\label{tab:cross_domain_performance}
\begin{tabular}{@{} l S[table-format=2.2] S[table-format=2.2] @{}}
\toprule
\textbf{Model Architecture} & \textbf{IEMOCAP (\%)} & \textbf{TESS (\%)} \\ 
\midrule
ResNet         & 41.60 & 20.86 \\
MobileNetV3    & 31.94 & \textbf{22.97} \\
ShuffleNetV2   & 34.13 & 21.33 \\
SqueezeNet     & \textbf{43.46} & 20.04 \\
GhostNet       & 33.40 & 21.15 \\
\bottomrule
\end{tabular}

\vspace{4pt}
\footnotesize \textit{Note}: All models trained exclusively on IEMOCAP. Accuracy values reflect in-domain (IEMOCAP) vs. cross-domain (TESS) generalization. Bold values indicate the best performance within each test scenario.
\end{table}
\subsection{Determination of Speech Feature Extraction Methods}
In the pre - experimental stage, we compared different speech feature extraction methods, including mel - spectrogram features, chromaticity features, and contrast features. We tested the TESS dataset and found that the accuracy of the mel - spectrogram feature on the test set reached 99.82\% (see Fig \ref{fig:trainsub1}), significantly better than the chromaticity feature (97.86\%) (see Fig \ref{fig:trainsub2}) and contrast feature (87.41\%) (see Fig \ref{fig:trainsub3}).  Based on these results, the mel - spectrogram feature showed stronger recognition ability and robustness, so we chose the mel - spectrogram as the optimal feature extraction method and widely used it in the training and evaluation of all models in subsequent experiments. We trained five models (ResNet, MobileNetV3, ShuffleNetV2, SqueezeNet, and GhostNet), all with a similar number of parameters. During training, mel-spectrogram features were used and trained on the IEMOCAP dataset (60\% training set, 40\% test set). Although every model converged to nearly 100\% training accuracy, their generalisation on the IEMOCAP test split varied considerably: 41.60\% for ResNet, 31.94\% for MobileNetV3, 34.13\% for ShuffleNetV2, 43.46\% for SqueezeNet, and 33.40\% for GhostNet (Table \ref{tab:cross_domain_performance}). The pronounced gap between training and test performance indicates severe over-fitting, motivating the introduction of a conformal-prediction framework to obtain more reliable model assessment and calibration in the remainder of this.

\begin{figure}[t]
  \centering
  \begin{subfigure}[b]{0.495\textwidth}
    \centering
    \includegraphics[width=\textwidth]{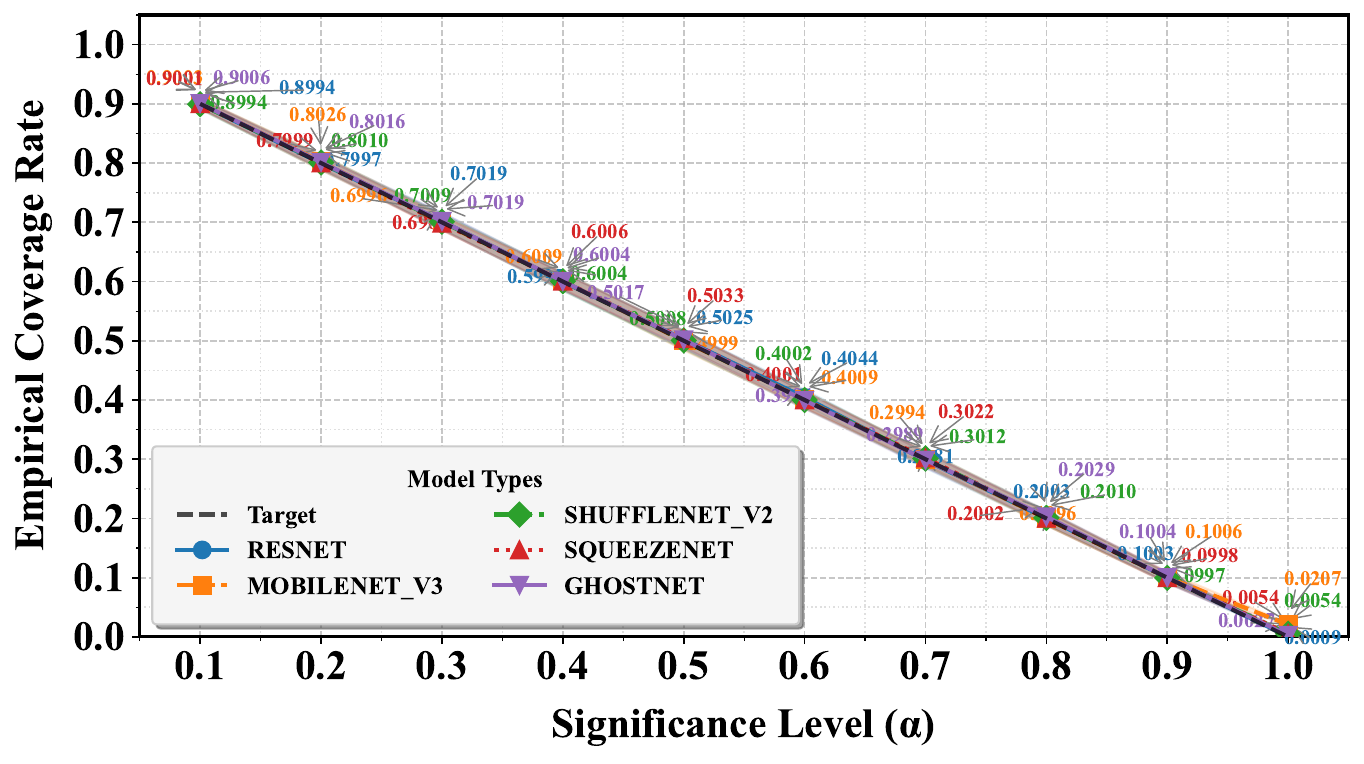}
    \caption{SCP framework on IEMOCAP}
    \label{fig:scp_iemocap}
  \end{subfigure}
  \hfill 
  \begin{subfigure}[b]{0.495\textwidth}
    \centering
    \includegraphics[width=\textwidth]{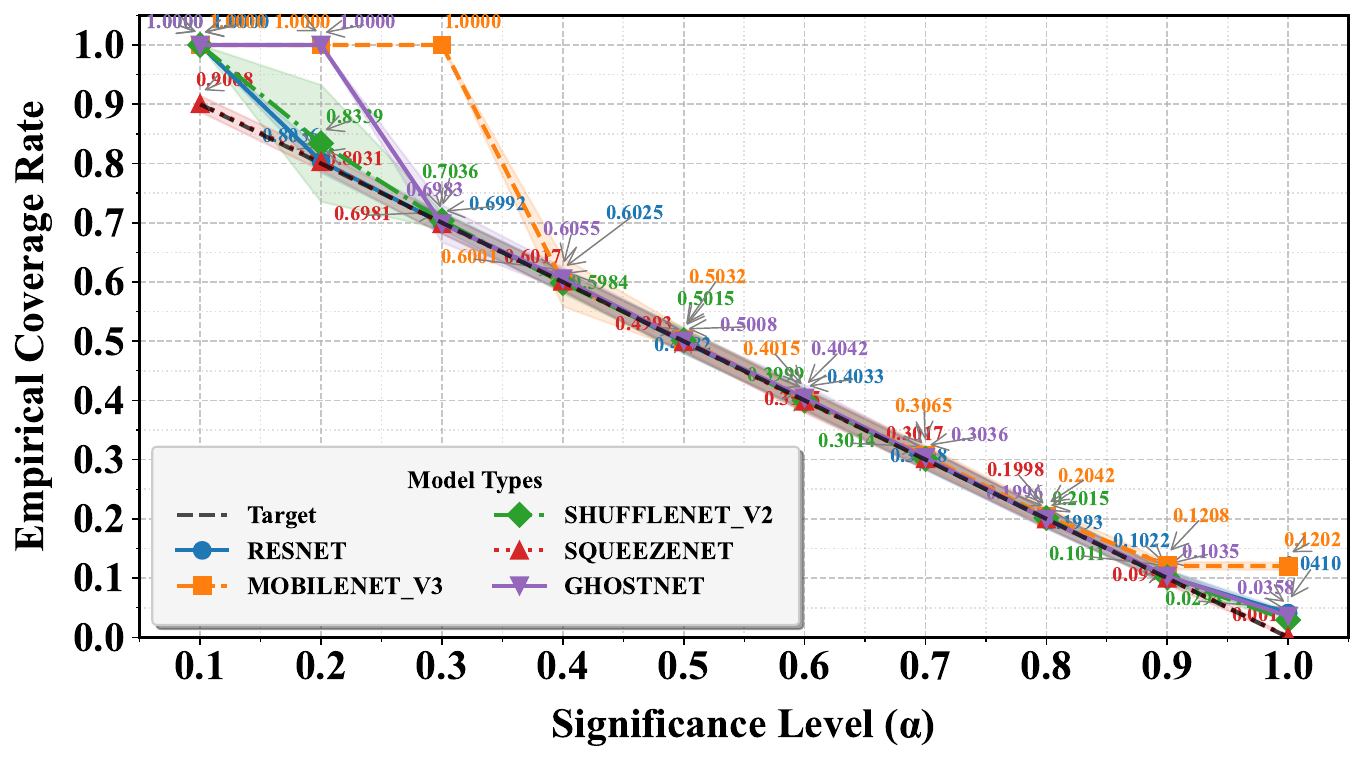}
    \caption{SCP framework on TESS}
    \label{fig:scp_tess}
  \end{subfigure}
  \caption{ECR using SCP framework}
  \label{fig:scp_group}
\end{figure}

\begin{figure}[t]
  \centering
  \begin{subfigure}[b]{0.495\textwidth}
    \centering
    \includegraphics[width=\textwidth]{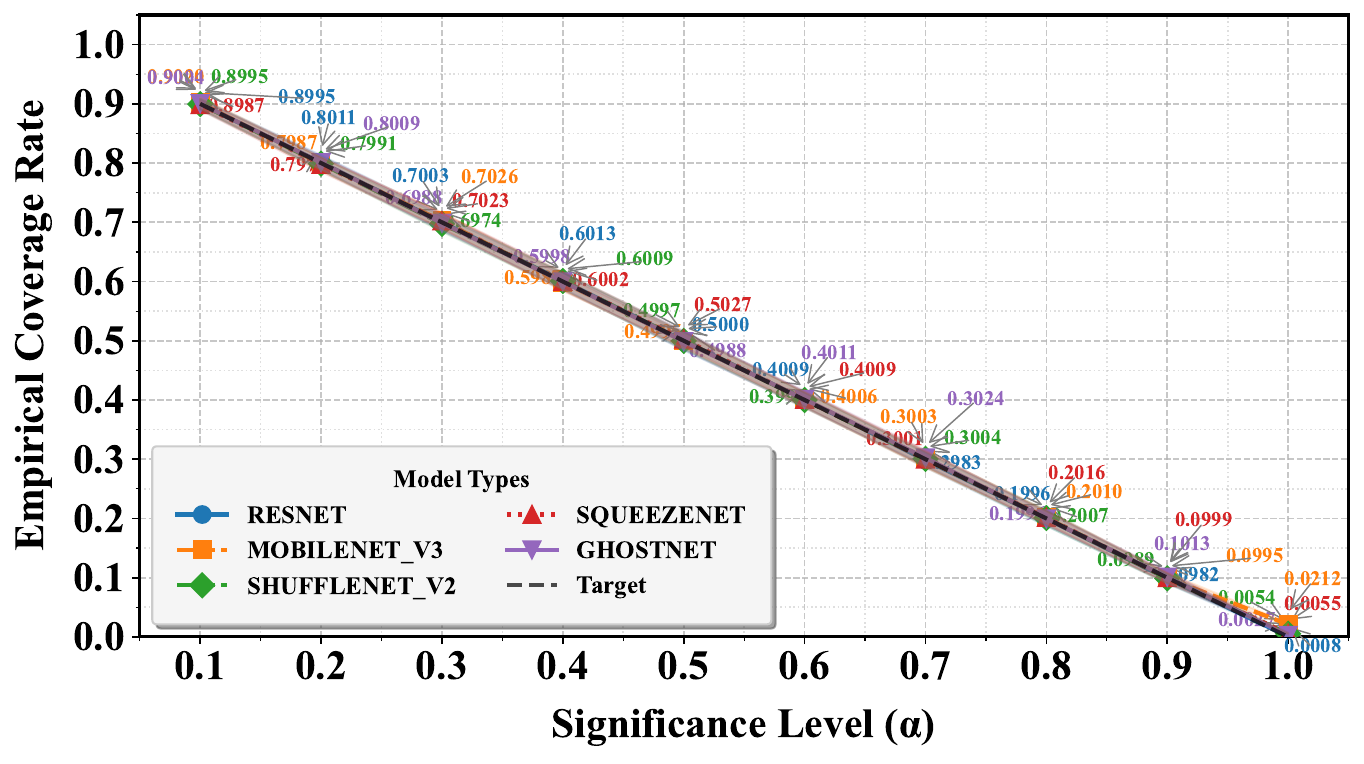}
    \caption{RCCP framework on IEMOCAP}
    \label{fig:risk_iemocap}
  \end{subfigure}
  \hfill
  \begin{subfigure}[b]{0.495\textwidth}
    \centering
    \includegraphics[width=\textwidth]{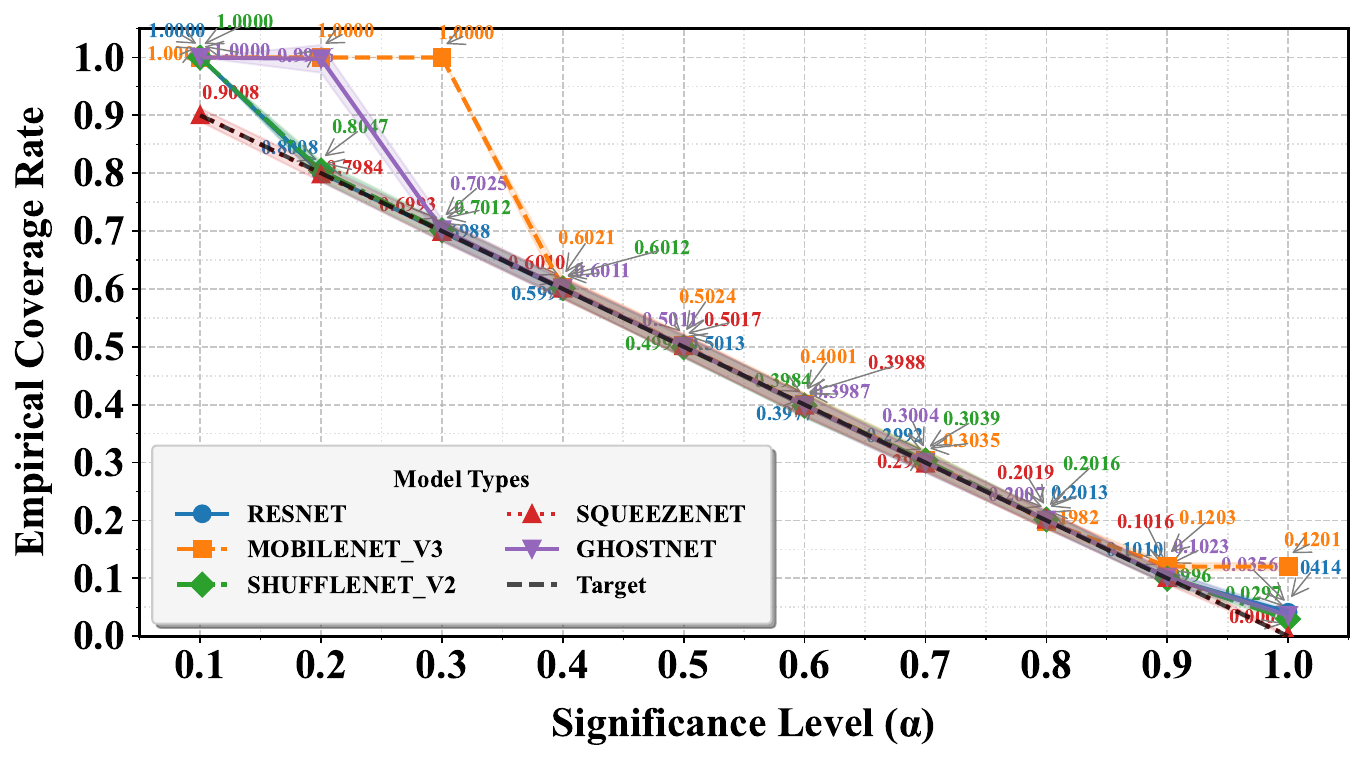}
    \caption{RCCP framework on TESS}
    \label{fig:risk_tess}
  \end{subfigure}
  \caption{ECR using RCCP framework}
  \label{fig:risk_group}
\end{figure}
\subsection{Empirical Results of the ECR metric}
We used SCP and RCCP to evaluate the coverage performance of five models on the IEMOCAP dataset, and applied these models to the TESS dataset for testing. On the IEMOCAP and TESS datasets, both frameworks provide reliable coverage guarantees on the IEMOCAP and TESS datasets. The experiment is repeated 100 times, and the calibration and test sets are resampled in each iteration, as shown in ~\cref{fig:scp_group,fig:risk_group} curves denote the Empirical Coverage Rate, and the shaded part indicates its standard deviation, the Empirical Coverage Rate of the five models is higher than the target coverage rate, and shows that both frameworks have good generalization ability on different datasets. While the theoretical guarantee of conformal prediction is rigorous, there can be minor fluctuations in practice due to finite - sample variability.

\subsection{Uncertainty Estimation of the Model through the APSS metric}
\begin{table}[t]
\centering
\footnotesize 
\setlength{\tabcolsep}{3pt}
\caption{The prediction efficiency of five IEMOCAP-trained models was quantified through Split Conformal Prediction and Risk-controlled Conformal Prediction, measuring average prediction set size (±SD) across two evaluation corpora (label counts specified parenthetically).}
\resizebox{\textwidth}{!}{ 
\begin{tabular}{@{}l l *{9}{c} @{}}
\toprule
\textbf{Dataset} & \textbf{Model} & \multicolumn{9}{c}{\textbf{Risk Level}} \\
\cmidrule(lr){1-2} \cmidrule(lr){3-11}
{} & {} & {0.1} & {0.2} & {0.3} & {0.4} & {0.5} & {0.6} & {0.7} & {0.8} & {0.9} \\
\midrule

\multicolumn{2}{@{}l@{} }{\textbf{Method: SCP}} & \multicolumn{9}{c}{} \\ 
\cmidrule{1-11}  

    \multirow{5}{*}{\centering IEMOCAP(6)} 
        & Resnet   & $4.025_{\pm0.049}$ & $3.106_{\pm0.050}$ & $2.319_{\pm0.044}$ & $1.755_{\pm0.040}$ & $1.297_{\pm0.030}$ & $0.941_{\pm0.025}$ & $0.621_{\pm0.023}$ & $0.362_{\pm0.019}$ & $0.160_{\pm0.011}$  \\
        & Mobilenetv3   & $4.570_{\pm0.062}$ & $3.705_{\pm0.063}$ & $2.948_{\pm0.052}$ & $2.346_{\pm0.064}$ & $1.818_{\pm0.040}$ & $1.344_{\pm0.035}$ & $0.922_{\pm0.026}$ & $0.529_{\pm0.027}$ & $0.217_{\pm0.015}$  \\
        & Shufflenetv2   & $4.468_{\pm0.065}$ & $3.534_{\pm0.048}$ & $2.800_{\pm0.053}$ & $2.236_{\pm0.047}$ & $1.720_{\pm0.041}$ & $1.252_{\pm0.028}$ & $0.832_{\pm0.032}$ & $0.494_{\pm0.022}$ & $0.211_{\pm0.016}$  \\
        & Squeezenet   & $3.975_{\pm0.072}$ & $2.989_{\pm0.051}$ & $2.230_{\pm0.048}$ & $1.659_{\pm0.044}$ & $1.237_{\pm0.031}$ & $0.868_{\pm0.021}$ & $0.576_{\pm0.025}$ & $0.342_{\pm0.017}$ & $0.147_{\pm0.011}$  \\
        & Ghostnet   & $4.657_{\pm0.064}$ & $3.698_{\pm0.065}$ & $2.973_{\pm0.052}$ & $2.319_{\pm0.052}$ & $1.761_{\pm0.037}$ & $1.276_{\pm0.028}$ & $0.855_{\pm0.027}$ & $0.519_{\pm0.020}$ & $0.235_{\pm0.016}$  \\
    \cmidrule{2-11}  
    \multirow{5}{*}{\centering TESS(4)}  
     & Resnet   & $4.000_{\pm0.000}$ & $2.893_{\pm0.059}$ & $2.551_{\pm0.035}$ & $2.230_{\pm0.041}$ & $1.894_{\pm0.044}$ & $1.580_{\pm0.040}$ & $1.225_{\pm0.029}$ & $0.663_{\pm0.120}$ & $0.191_{\pm0.017}$  \\
    & Mobilenetv3   & $4.000_{\pm0.000}$ & $4.000_{\pm0.000}$ & $4.000_{\pm0.000}$ & $2.210_{\pm0.182}$ & $1.846_{\pm0.050}$ & $1.526_{\pm0.039}$ & $1.161_{\pm0.047}$ & $0.662_{\pm0.056}$ & $0.281_{\pm0.008}$  \\
    & Shufflenetv2   & $4.000_{\pm0.000}$ & $3.106_{\pm0.558}$ & $2.490_{\pm0.051}$ & $2.161_{\pm0.047}$ & $1.819_{\pm0.042}$ & $1.480_{\pm0.038}$ & $1.125_{\pm0.034}$ & $0.626_{\pm0.047}$ & $0.199_{\pm0.022}$  \\
    & Squeezenet   & $3.621_{\pm0.051}$ & $3.190_{\pm0.055}$ & $2.751_{\pm0.060}$ & $2.364_{\pm0.064}$ & $1.897_{\pm0.056}$ & $1.483_{\pm0.059}$ & $0.977_{\pm0.045}$ & $0.535_{\pm0.042}$ & $0.222_{\pm0.018}$  \\
    & Ghostnet   & $4.000_{\pm0.000}$ & $4.000_{\pm0.000}$ & $2.439_{\pm0.026}$ & $2.176_{\pm0.043}$ & $1.873_{\pm0.040}$ & $1.562_{\pm0.052}$ & $1.216_{\pm0.035}$ & $0.695_{\pm0.059}$ & $0.255_{\pm0.024}$  \\
\midrule

\multicolumn{2}{@{}l@{} }{\textbf{Method: RCCP}} & \multicolumn{9}{c}{} \\ 
\cmidrule{1-11}  
    \multirow{5}{*}{IEMOCAP(6)} 
        & Resnet   & $4.017_{\pm0.061}$ & $3.091_{\pm0.059}$ & $2.306_{\pm0.041}$ & $1.759_{\pm0.042}$ & $1.294_{\pm0.029}$ & $0.935_{\pm0.027}$ & $0.619_{\pm0.022}$ & $0.361_{\pm0.017}$ & $0.155_{\pm0.012}$ \\
        & Mobilenetv3   & $4.574_{\pm0.060}$ & $3.691_{\pm0.058}$ & $2.951_{\pm0.049}$ & $2.333_{\pm0.043}$ & $1.828_{\pm0.042}$ & $1.340_{\pm0.042}$ & $0.922_{\pm0.027}$ & $0.532_{\pm0.025}$ & $0.215_{\pm0.015}$  \\
        & Shufflenetv2   & $4.464_{\pm0.064}$ & $3.553_{\pm0.054}$ & $2.801_{\pm0.054}$ & $2.230_{\pm0.048}$ & $1.713_{\pm0.045}$ & $1.240_{\pm0.032}$ & $0.823_{\pm0.029}$ & $0.495_{\pm0.024}$ & $0.210_{\pm0.016}$ \\
        & Squeezenet   & $3.964_{\pm0.061}$ & $2.979_{\pm0.062}$ & $2.217_{\pm0.048}$ & $1.666_{\pm0.037}$ & $1.234_{\pm0.032}$ & $0.868_{\pm0.026}$ & $0.569_{\pm0.025}$ & $0.344_{\pm0.018}$ & $0.149_{\pm0.013}$  \\
        & Ghostnet   & $4.649_{\pm0.059}$ & $3.707_{\pm0.071}$ & $2.979_{\pm0.056}$ & $2.313_{\pm0.049}$ & $1.764_{\pm0.037}$ & $1.271_{\pm0.033}$ & $0.859_{\pm0.030}$ & $0.518_{\pm0.021}$ & $0.235_{\pm0.015}$  \\
    \cmidrule{2-11}  
    \multirow{5}{*}{TESS(4)} 
      & Resnet  & $4.000_{\pm0.000}$ & $2.887_{\pm0.042}$ & $2.555_{\pm0.042}$ & $2.219_{\pm0.040}$ & $1.904_{\pm0.039}$ & $1.581_{\pm0.043}$ & $1.223_{\pm0.037}$ & $0.666_{\pm0.115}$ & $0.188_{\pm0.017}$  \\
    & Mobilenetv3   & $4.000_{\pm0.000}$ & $4.000_{\pm0.000}$ & $4.000_{\pm0.000}$ & $2.214_{\pm0.051}$ & $1.849_{\pm0.041}$ & $1.513_{\pm0.044}$ & $1.158_{\pm0.043}$ & $0.649_{\pm0.061}$ & $0.280_{\pm0.009}$  \\
    & Shufflenetv2   & $4.000_{\pm0.000}$ & $2.840_{\pm0.047}$ & $2.494_{\pm0.048}$ & $2.153_{\pm0.039}$ & $1.814_{\pm0.040}$ & $1.477_{\pm0.039}$ & $1.126_{\pm0.033}$ & $0.611_{\pm0.044}$ & $0.196_{\pm0.023}$  \\
    & Squeezenet   & $3.613_{\pm0.046}$ & $3.192_{\pm0.051}$ & $2.751_{\pm0.052}$ & $2.359_{\pm0.058}$ & $1.921_{\pm0.065}$ & $1.477_{\pm0.063}$ & $0.978_{\pm0.042}$ & $0.530_{\pm0.033}$ & $0.221_{\pm0.021}$ \\
    & Ghostnet   & $4.000_{\pm0.000}$ & $3.975_{\pm0.174}$ & $2.486_{\pm0.051}$ & $2.174_{\pm0.050}$ & $1.865_{\pm0.046}$ & $1.556_{\pm0.052}$ & $1.215_{\pm0.031}$ & $0.676_{\pm0.065}$ & $0.249_{\pm0.024}$  \\
\bottomrule
\end{tabular}
}

\label{tab:framework_comparison}
\end{table}
We found a negative correlation between the average prediction set size (APSS) and the risk level. Specifically, as shown in the table \ref{tab:framework_comparison}, the APSS value decreases significantly as the risk level increases,, indicating that the model's predictions are more certain at higher risk levels and the size of the prediction set is optimized. In this way, APSS is able to effectively evaluate the uncertainty of the classification model on the test set. Larger APSS values indicate that the model has higher prediction uncertainty for certain categories, while smaller APSS values mean that the model has higher confidence in its predictions. Therefore, APSS, as a measure of model uncertainty, demonstrates its potential as a promising benchmark in evaluating the robustness of classification models and provides a valuable reference for further optimizing models.

\subsection{Ablation Studies}
\begin{figure}[t]
    \centering
    \includegraphics[width=1.0\textwidth]{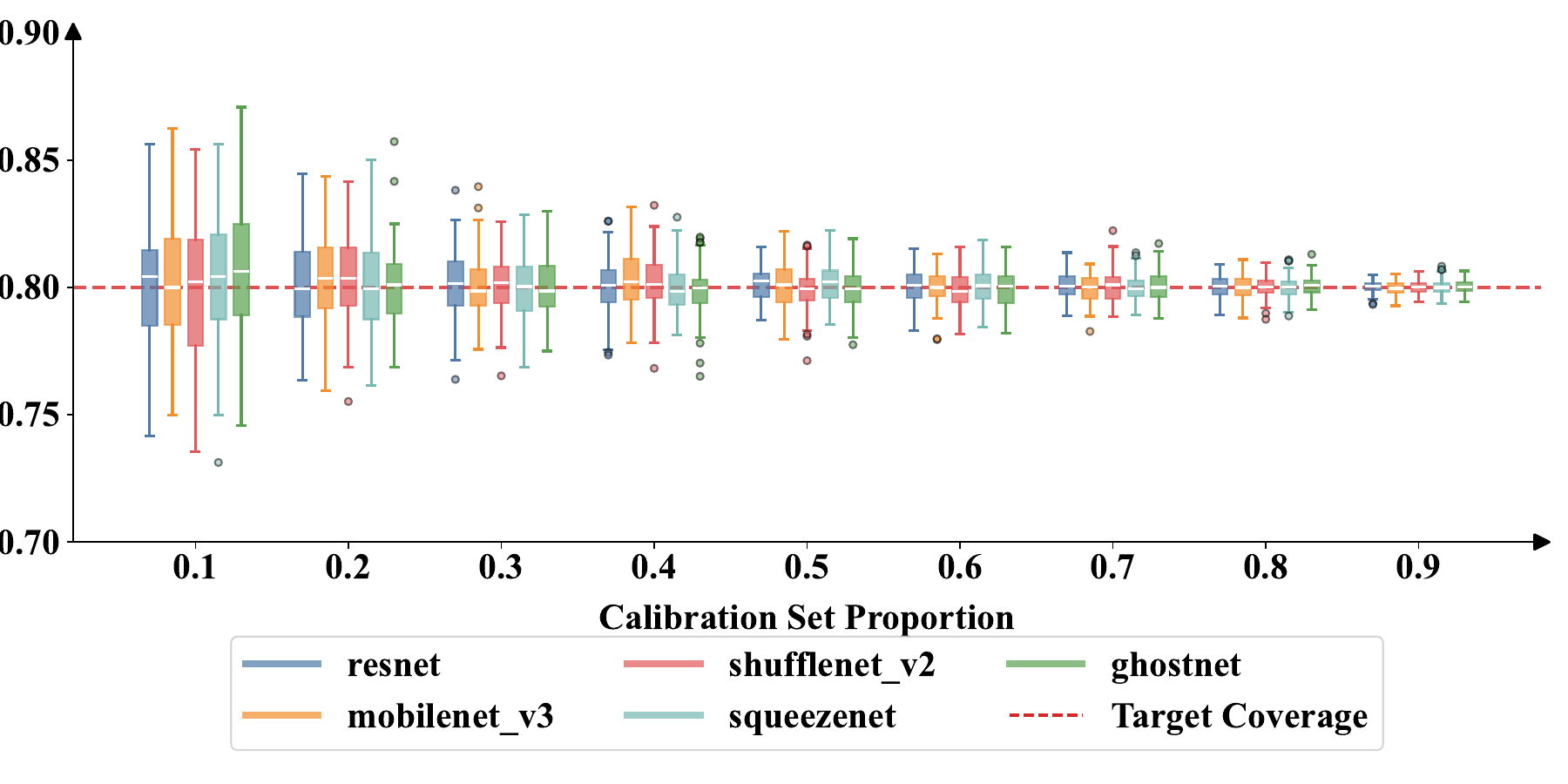} 
    \caption{Coverage Distribution vs Calibration Set Size ($\alpha = 0.2$)}
    \label{fig_xiaorong}
\end{figure}
This study systematically explores the impact of different data partitioning strategies on model coverage and verifies the robustness of the speech emotion recognition method based on the conformal prediction framework. In the experimental design, we adopt a dynamic calibration mechanism: the confidence threshold $\hat{q}$ is calculated through the calibration set data, and its statistical validity is verified under different calibration set - test set partition ratios (10\%--90\%).
\begin{figure}[h]
    \centering
    \begin{subfigure}[b]{0.9\textwidth}
        \centering
        \includegraphics[width=\textwidth]{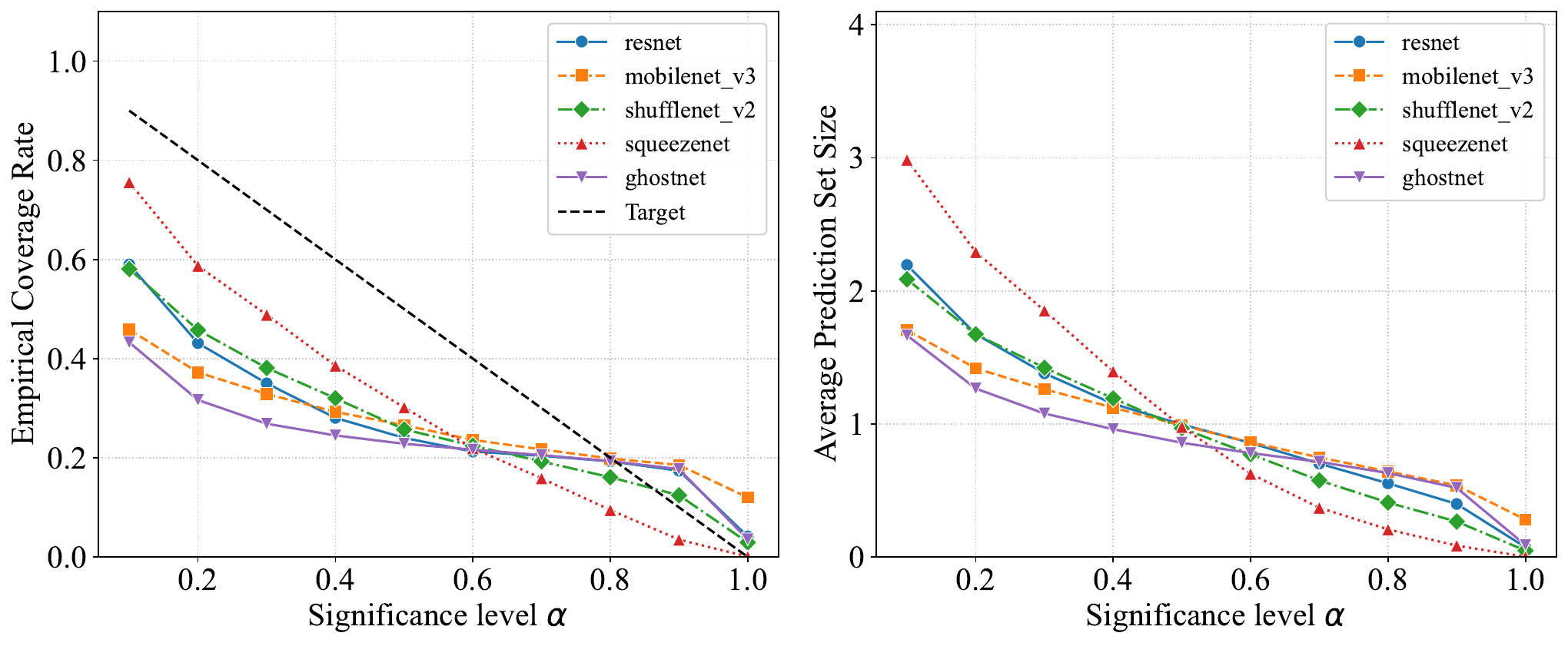}
        \caption{SCP method}
        \label{fig:image1}
    \end{subfigure}
    \vspace{0.1cm} 
    \begin{subfigure}[b]{0.9\textwidth}
        \centering
        \includegraphics[width=\textwidth]{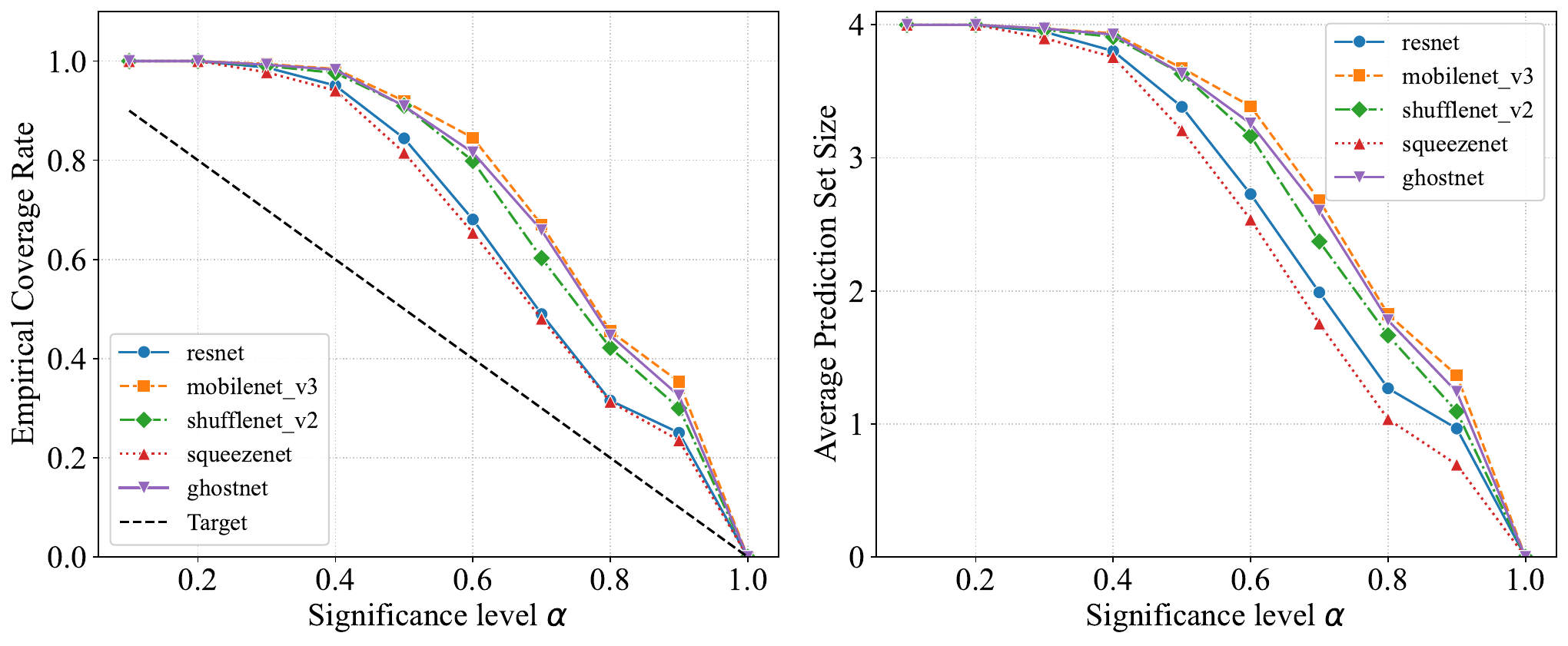}
        \caption{martingale method}
        \label{fig:image2}
    \end{subfigure}
    \caption{Coverage guarantee based on martingale method}
    \label{fig:yangzhi}
\end{figure}
The experimental results show that although the calibration set size changes significantly (from 10\% to 90\%), this method always ensures the empirical coverage of the test set satisfies the theoretical lower bound $1-\alpha$ ($\alpha = 0.2$), as shown in the figure \ref{fig_xiaorong} by dynamically adjusting the confidence threshold $\hat{q}$. Theoretical analysis and empirical results jointly prove that this method exhibits strong robustness to data partition ratios. The constructed prediction sets provide statistically rigorous reliability guarantees for speech emotion recognition tasks.

\subsection{Relax the exchangeability based on small batch martingale process}
In order to verify the effectiveness of the proposed martingale method and the rationality of the assumption of local interchangeability in small batch scenarios, this paper adopts a cross-dataset experimental design, focusing on the actual scenario where inter-batch data does not meet global interchangeability. The calibration set is selected from the IEMOCAP speech emotion dataset, and the test set is the TESS dataset. There are significant differences in the speaker, emotional expression mode and acoustic environment which directly leads to inconsistent data distribution between batches (that is, the inter-batch does not meet the interchangeability). The experiment divides the data into "local batches", each batch contains 5 calibration data and 1 test data, strengthening the local correlation of internal data by limiting the batch size - although there may be significant distribution differences between batches (such as grouping characteristics of different time periods and different scenarios), the data within a single batch can approximately meet "local interchangeability" because it comes from a closer local distribution.
\begin{figure}[t] 
  \centering 
  \includegraphics[width=1\textwidth]{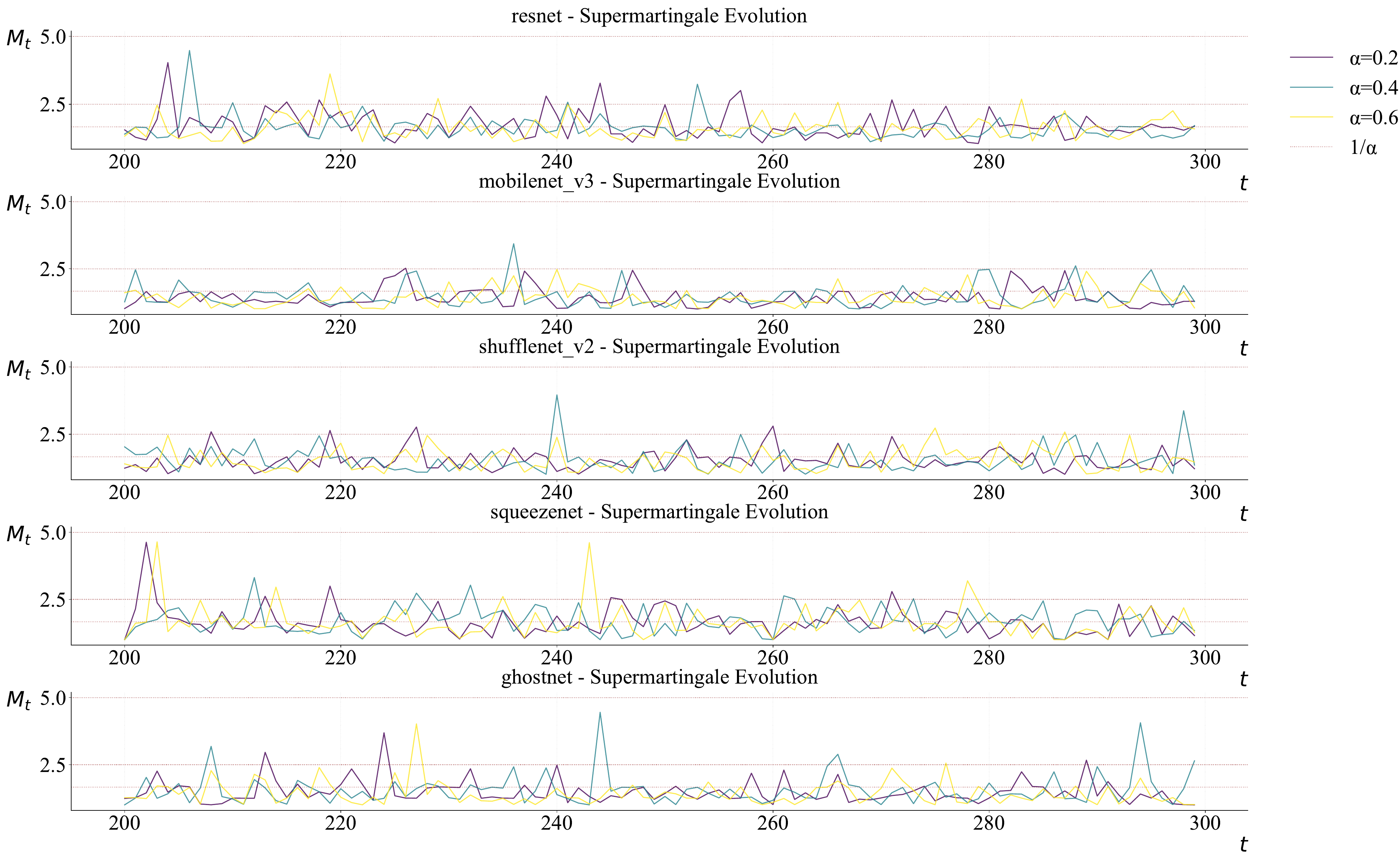} 
  \caption{Supermartingale Analysis} 
  \label{fig:yang} 
\end{figure}
As shown in the figure \ref{fig:yangzhi}, traditional methods experience significant coverage bias when data does not satisfy exchangeability, as the global exchangeability assumption fails, resulting in empirical coverage below the target value. This method uses the local exchangeability assumption and martingale value screening, and when the distribution difference between batches is significant, the empirical coverage rate and target value are still achieved, verifying the validity of the relaxation assumption that "interchangeability can not be met between batches".$M_{t}$ significantly affects the quality of the conformal set at batch t. As the martingale approaches zero, the conformal set becomes larger and less informative. The figure \ref{fig:yang} shows the changing path of $M_{t}$. It is worth noting that this method ensures coverage but does not control the size of the conformal set; stronger models generally produce smaller sets.

This framework does not require global data to obey the same distribution, significantly relaxing the prerequisites for traditional conformal prediction, and is suitable for complex scenarios where there is a distribution offset between batches.

\section{Conclusion}
This study proposes a solution based on SCP and RCCP framework to solve the problem of unreliable prediction of traditional neural networks in speech emotion recognition. The results show that traditional point prediction methods have significant confidence bias due to factors such as data distribution differences, model overfitting, and difficulty in capturing complex emotional features. By using the SCP framework and RCCP framework, this study successfully transforms single-point predictions into set predictions with coverage guarantees, ensuring that the true label is included in the prediction set with a probability of at least $1-\alpha$ under a specified risk level $\alpha$. In addition, cross-dataset validation proves the robustness of the framework under different data distributions.
The study also finds that APSS is negatively correlated with the risk level, indicating that when the risk level is high, the model's prediction set is usually smaller and the model's prediction uncertainty is lower. Based on this finding, we propose a metric for evaluating the uncertainty of classification models on the test set. At the same time, our mini-batch martingale value conformal prediction theory extends the applicability of CP to non-exchangeable environments, solving a key flaw in existing uncertainty quantification methods. By adopting the local exchangeability assumption within the mini-batch and constructing a non-negative test martingale, our method maintains coverage guarantees even under non-exchangeable data.

\bibliographystyle{unsrt}
\bibliography{reference}  






\end{document}